\documentclass[epj]{svjour}
\usepackage[utf8]{inputenc}
\usepackage[normalem]{ulem}
\usepackage{amsmath}
\usepackage{amssymb}
\usepackage{MnSymbol}
\usepackage{bbold}
\usepackage{graphicx}
\usepackage{braket}
\usepackage[dvipsnames]{xcolor}

\usepackage{orcidlink}
\hypersetup{
    colorlinks=true,       
    linkcolor=blue,          
    citecolor=blue,        
    filecolor=blue,      
    urlcolor=blue           
}

\begin{document}
%


%
\title{
\includegraphics[scale=0.1]{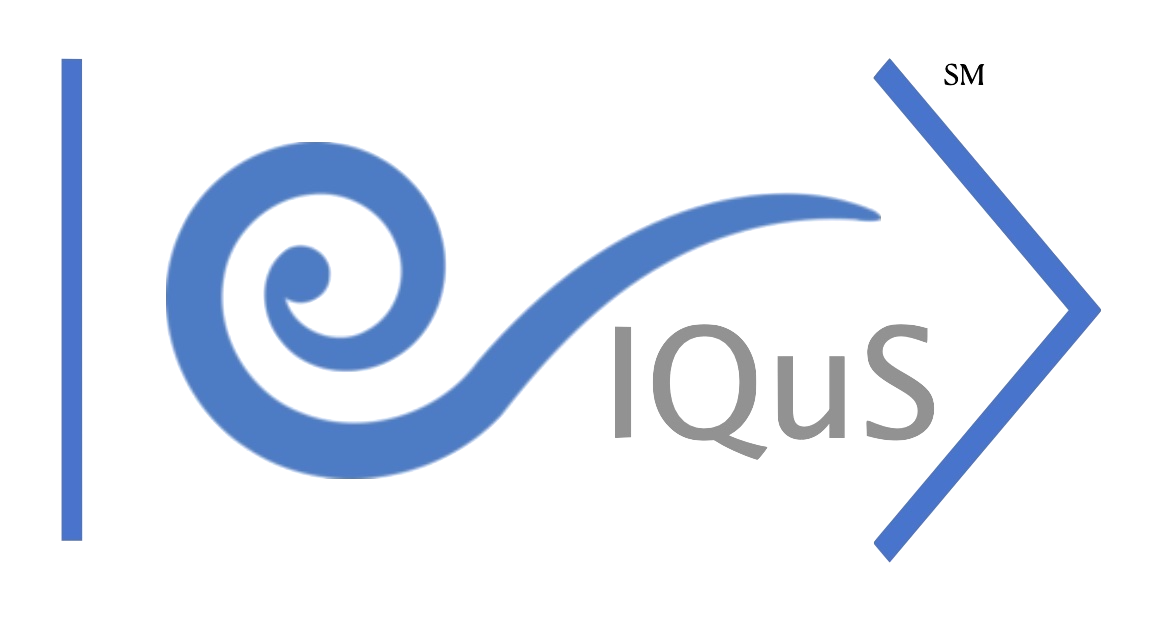} \\  {\small IQuS@UW-21-053} \\ \\ \\Multi-Body Entanglement and Information Rearrangement in Nuclear Many-Body Systems
}



\author{S. Momme Hengstenberg\inst{1}\thanks{shengstenberg@physik.uni-bielefeld.de}\orcidlink{0009-0009-8764-9980}
\and Caroline E. P. Robin\inst{1,2} \thanks{crobin@physik.uni-bielefeld.de} \orcidlink{0000-0001-5487-270X} 
\and Martin J. Savage\inst{3}\thanks{mjs5@uw.edu . On leave from the Institute for Nuclear Theory.} \orcidlink{0000-0001-6502-7106}}

\institute{Fakult\"at f\"ur Physik, Universit\"at Bielefeld, D-33615, Bielefeld, Germany. 
\and GSI Helmholtzzentrum f\"ur Schwerionenforschung, Planckstra{\ss}e 1, 64291 Darmstadt, Germany. 
\and InQubator for Quantum Simulation (IQuS), Department of Physics, University of Washington, Seattle, WA 98195, USA.   }

\abstract{
We examine how effective-model-space (EMS) calculations of nuclear many-body systems rearrange and converge multi-particle entanglement.
The generalized Lipkin-Meshkov-Glick (LMG) model is used to motivate and provide insight for future developments of entanglement-driven descriptions of nuclei.
The effective approach is based on a truncation of the Hilbert space together with a variational rotation of the qubits (spins), which constitute the relevant elementary degrees of freedom. The non-commutivity of the rotation and truncation allows for an exponential improvement of the energy convergence throughout much of the model space.
Our analysis examines measures of correlations and entanglement, and quantifies their convergence with increasing cut-off. We focus on one- and two-spin entanglement entropies, mutual information, and $n$-tangles for $n=2,4$ to estimate multi-body entanglement. The effective description strongly suppresses entropies and mutual information of the rotated spins, while being able to recover the exact results to a large extent with low cut-offs. Naive truncations of the bare Hamiltonian, on the other hand, artificially underestimate these measures. The $n$-tangles in the present model provide a basis-independent measures of $n$-particle entanglement. 
While these are more difficult to capture with the EMS description, the improvement in convergence, compared to truncations of the bare Hamiltonian, is significantly more dramatic. We conclude that the low-energy EMS techniques, that successfully provide predictive capabilities for low-lying observables in many-body systems,
exhibit analogous efficacy for quantum correlations and multi-body entanglement in the LMG model, motivating future studies in nuclear many-body systems and effective field theories relevant to high-energy physics and nuclear physics.
}


%
\maketitle

%
\section{Introduction}
\label{sec:intro}
\noindent
Entanglement is a common feature of interacting quantum many-body systems. 
This phenomenon arises due to the interaction between the constituents of a system, making them correlated in a non-local manner, and thus strongly influences both the structure and dynamics of these systems.
Because they are entangled, interacting particles cannot be described independently from each others, and instead have to be considered as a whole.
In the context of solving the many-body problem, this means that the state of the system cannot be represented by a simple tensor product state, or classical configuration, and instead is an entangled superposition of those configurations, whose number grows {\it a priori} exponentially with the number of degrees of freedom. Such exponential scalings make quantum many-body problems typically hard to solve on classical computers, as they require exponential amounts of computational resources (time and/or memory).

On the other hand, it is well known that the number of relevant states needed to describe low-energy eigenstates of many interesting systems is typically small compared to the size of the full of the Hilbert space.
In the context of classical calculations, finding which configurations are the most relevant, and being able to group them within a "small corner" of the Hilbert space allows one to make truncations and render the problem more tractable with limited loss of information. Such concepts are implemented by methods such as density matrix renormalization group (DMRG)~\cite{PhysRevLett.69.2863,SCHOLLWOCK201196} and tensor networks~\cite{Or_s_2019}, that use 
singular value decomposition (SVD) of the reduced density matrix (given some bi-partition of the system) to compress and truncate the model space. These methods however rely on the fact that 
entanglement is organized such that a truncation in the singular value spectrum can be safely applied~\cite{RevModPhys.82.277,PhysRevLett.91.147902}. This makes them particularly powerful for the description of systems satisfying entanglement area laws, such as low-lying states of many physically relevant quantum lattice systems, for which the entanglement entropy of a subpart of the system grows only proportionally with its boundary area~\cite{RevModPhys.82.277}. 

Entanglement entropies in atomic nuclei are expected to satisfy volume laws~\cite{Pazy:2022mmg,Gu:2023aoc}, and thus to be {\it a priori} distributed over many components. 
Nevertheless attempts at applying DMRG, or similar methods, in these systems have been made successful in certain nuclei presenting a natural weakly entangled bi-partioning. 
This includes, for example, nuclear wave function factorization based on a bi-partition in terms of proton and neutron subsectors \cite{Papenbrock_2005,Johnson:2022mzk} which has proven to be efficient in nuclei away from $N=Z$. 
Versions of DMRG that resemble White's original formulation~\cite{PhysRevLett.69.2863} have been found successful for the description of light weakly-bound nuclei because of weak entanglement between valence particles in the continuum and the core of the nucleus \cite{PhysRevLett.97.110603,PhysRevC.79.014304,PhysRevC.88.044318}. Such calculations however rapidly become computationally challenging when going towards medium-mass nuclei~\cite{PhysRevC.65.054319,Dukelsky:2004vv,PhysRevC.78.041303,PhysRevC.106.034312}.

In systems where no bi-partitioning is evident, being able to manipulate, transform and decrease the amount of entanglement can potentially be useful in order to make such approaches more efficient. 
In a previous study, we have explored the possibility of rearranging and minimizing entanglement structures of nuclei into localized regions of the Hilbert space via bases transformations~\cite{Robin:2020aeh}. 
More precisely, we have investigated the entanglement patterns emerging from {\it ab-initio} EMS calculations of light Helium nuclei. We studied how these structures can be rearranged via unitary transformations of the effective Hamiltonian corresponding to a rotation of the orbital basis, obtained from a variational principle. Calculations of entanglement entropies, mutual information and negativity showed that such transformations, which accelerate the convergence of the wavefunction and binding energy, are able to minimize and move entanglement into the effective model space in a natural way. 
These transformations also led to an emergent picture of $^6$He consisting of two interacting valence neutrons decoupling from an $^4$He core, thus driving the wave function to a core-valence tensor-product structure, and revealing new degrees of freedom.
As we mentioned in that study~\cite{Robin:2020aeh}, combining this method with a DMRG framework 
is part of a future work that is underway. 
Other attempts at using entanglement rearrangement for improving the computation efficiency include the application of
DMRG guided by tools of quantum information, as initially applied in quantum chemistry~\cite{PhysRevB.68.195116}. 
Such methods have been adapted to nuclei in the context of phenomenological shell-model calculations \cite{PhysRevC.92.051303}, and more recently in an {\it ab-initio} framework~\cite{Tichai:2022bxr}. 
In those studies,
the orbitals were re-ordered based on their mutual information without variational transformations, and thus without modification of the total entanglement content.
\\
\\ 
\indent 
From a more fundamental perspective, 
it has also been realized that formal concepts of quantum information, in particular those related to the characterization of entanglement, may provide new ways of understanding the structure of matter and interaction between constituents, and thus offer new possibilities in formulating theories for the description of nature.
In the hadronic sector, it has been shown that entanglement suppression is related to the emergence of symmetries of the strong interaction at low energies~\cite{Beane:2018oxh,Beane:2020wjl,Beane:2021xrk,Low:2021ufv,Liu:2022grf},
which are observed in lattice-QCD calculations of two-baryon systems~\cite{NPLQCD:2012mex,Wagman:2017tmp},
and manifested in the structure and dynamics of nuclei and their reactions, 
even beyond the constraints imposed by the large-$N_c$ limit of quantum chromodynamics (QCD)~\cite{Kaplan:1995yg}.
Relations between entanglement and few-nucleon scatterings have also recently been investigated in Refs.~\cite{Bai:2022hfv,Bai:2023tey,Miller:2023ujx,Miller:2023snw}.
Although investigations of entanglement in many-body nuclear systems have overall started more recently than in other fields of quantum many-body physics, the development of quantum computers and quantum information science has triggered increasing efforts in this area. 
Studies of entanglement in nuclei include investigations of entanglement between proton and neutron sub-systems \cite{PhysRevC.67.051303,PhysRevC.69.024312,Johnson:2022mzk}, as well as investigation of mode or orbital entanglement within phenomenological shell-model-type calculations \cite{PhysRevC.92.051303,Kruppa_2021,PhysRevC.106.024303,PhysRevC.105.064308,Perez-Obiol:2023vod} and {\it ab-initio} calculations \cite{Robin:2020aeh,Tichai:2022bxr}. Entanglement induced by short-range correlations has been recently studied in Refs.~\cite{Bulgac:2022cjg,Bulgac:2022ygo,Pazy:2022mmg} while entanglement and correlations of nuclear models have been investigated in Refs.~\cite{PhysRevA.103.032426,PhysRevA.104.032428,PhysRevA.105.062449,Gu:2023aoc}.
\\
\\ 
\indent 
Understanding and manipulating entanglement is of course also particularly relevant to the development of quantum simulations of many-body systems, in particular in the context of NISQ devices, that are subject to noise, limited number of qubits and/or connectivity, imperfect gates and measurements, and thus require short, truncated quantum circuits. In that context, being able to rearrange entanglement into localized structures 
({\it i.e.} into a small number of qubits) can be highly beneficial, allowing for efficient use of quantum resources 
and reduction of the uncertainties associated with simulations.
In addition, in hybrid high-performance computing (HPC) and quantum computing 
architectures, the ability to rearrange entanglement is an important ingredient in
making optimal use of such a heterogeneous environment~\cite{BDKS:2023cool}.
For nuclear physics and high-energy physics applications, such optimization(s) are in early stages.

In a broader context, quantum many-body systems play a central role in 
detecting, manipulating and storing quantum information.  
For example, in the area of quantum sensing, 
sensitivity below the standard quantum limit can be achieved through 
creating entanglement structures in many-body systems, such as  the
spin-squeezing of multiple 
spins~\cite{PhysRevA.47.5138,PhysRevA.46.R6797,PhysRevA.50.67,PhysRevA.68.012101}. 
Such states can be created using unitary operators similar to, for example,
terms in the LMG model Hamiltonian.
In quantum computing, 
creating entanglement structures in quantum many-body systems with particular properties 
is at the heart of 
error correction, including the form of logical qubits comprised of noisier physical qubits.

A number of works have now developed and/or adapted quantum algorithms for simulations of nuclei and nuclear systems. 
These include applications of the variational quantum eigensolver (VQE), or variants thereof, for the determination of ground- and excited-states properties of nuclear systems including the LMG model \cite{PhysRevC.104.024305,Hlatshwayo:2022yqt,Chikaoka:2022kff,PhysRevC.105.064317,Robin:2023pgi}, light nuclei \cite{PhysRevLett.120.210501,PhysRevA.100.012320,PhysRevC.105.064308,PhysRevC.106.034325}, as well as medium-mass nuclei in small valence spaces \cite{PhysRevC.105.064317,Perez-Obiol:2023vod}. 
Algorithms for breaking and restoring symmetries have also been developed and applied to the Richardson pairing model in Refs.~\cite{PhysRevC.105.024324,Lacroix:2022vmg,PhysRevC.107.034310}.
Algorithms for describing nuclear dynamics have been developed and applied to the Agassi model \cite{P_rez_Fern_ndez_2022,Saiz:2022rof,Illa:2023scc}, as well as the dynamics and response of toy few-nucleon systems \cite{PhysRevA.101.062307,PhysRevC.102.064624,Turro:2023xgf,PhysRevD.101.074038,PhysRevD.105.074503}.
Of relevance for quantum simulations of observables within a specific energy interval,
as commonly required in nuclear physics and high-energy physics processes, are 
algorithms implementing a band-pass filter, such as those in Refs.~\cite{Lu_2021,Choi:2020pdg}.

So far, however, utilizing entanglement as ingredient to guide the simulations 
and improve their efficiency has received limited attention.
Recently, we have taken a step in this direction~\cite{Robin:2023pgi}
and designed a hybrid classical-quantum algorithm, Hamiltonian learning variational quantum eigensolver (HL-VQE), that simultaneously optimizes an effective Hamiltonian, thereby rearranging entanglement into the effective model space, and the associated ground-state wavefunction.
We applied this procedure to the LMG model and found that HL-VQE provides an exponential improvement in the convergence of the ground-state energy and wavefunction fidelity, as compared to VQE alone (i.e. without optimizing the effective Hamiltonian), throughout a large part of the Hilbert space. Implementations on IBM’s QExperience~\cite{IBMQ} quantum computers and simulators were able to reproduce classical predictions precisely and accurately, with only a few qubits.

In the present paper, we investigate an EMS framework
from the point of view of entanglement, to better understand the mechanisms for improving both classical and quantum computations of nuclear systems.
We consider the generalized version of the LMG model, introduced in Ref.~\cite{LIPKIN1965188}, to provide a toy model presenting generic features of atomic nuclei such as symmetry-breaking and phase transitions.
In this model, due to the nature of the interactions, the relevant elementary degrees of freedom are two-level quantum systems, {\it i.e.} qubits. For this reason the LMG model can also be mapped into a system of spin-$1/2$ particles and has been the subject of numerous studies in the context of quantum information (see, for example, Refs.~\cite{PhysRevA.68.012101,PhysRevA.69.022107,PhysRevA.69.054101,PhysRevA.70.062304,PhysRevA.71.064101,PhysRevA.77.052105,PhysRevLett.101.025701,PhysRevA.100.062104,PhysRevB.101.054431,Calixto_2021,PhysRevA.103.032426,PhysRevA.104.032428,PhysRevA.105.062449}). These studies typically focused on exact or mean-field calculations, and relations between entanglement measures and phase transitions.
We emphasize that the focus of the present study is the EMS method and its impact on entanglement, rather than the LMG model itself, which is used as a ``sandbox''. The aim is to gain insight for future studies of realistic nuclear systems.
In the case of the LMG model, the EMS method is applied by introducing a truncated EMS determined by single-particle excitations, or equivalently by the number of spins up, and by performing variational rotations of the qubits, or spins, thereby defining an effective Hamiltonian for the EMS. The rotated spins provide effective degrees of freedom which are expected to exhibit reduced correlations and entanglement with others, while capturing, in part, the physics of the fully interacting system solved in an exact way. Such aspects are extensively investigated in the present study.

The paper is organized as follows.
In Sec.~\ref{sec:LMG}, the reader is reminded about relevant
aspects of the LMG model, and how the EMS framework applies to this model. 
The convergence of the energy with increasing 
dimensionality of the EMS is analyzed, 
and comparisons with calculations that apply a 
naive truncation of the bare Hamiltonian ({\it i.e.}, without variational rotations of the spins), are provided.
In Sec.~\ref{sec:entang_LMG}, we examine various measures of correlations and entanglement, which probe different aspects of the many-body wave function, and study their convergence behaviours. 
Specifically, one- and two-spin entanglement entropies are calculated, as well as mutual information, to investigate how the bare and rotated (effective) spins are entangled within the system. 
We also examine the $n$-tangles, $\tau_n$, 
which provide a measure of $n$-body entanglement, and which 
have not been previously studied in the LMG model (except for the case $n=2$). 
In particular, the 2- and 4-tangles are determined, 
and analytical expressions for $\tau_n$ are presented up to $n=6$ . 
The $n$-tangles, which we find to be basis-independent in the LMG model, probe 
fine details of the many-body wave function and thus provide a stringent test of the many-body method. 
Convergence studies of these measures can thus be very informative. 
We perform such studies of $\tau_2$ and $\tau_4$,
and compare to the naively truncated calculations using the bare Hamiltonian.
In Sec.~\ref{sec:spin_sq_LMG}, spin-squeezing within the LMG model is evaluated using  
results of quantum simulations that
we previously performed using IBM’s quantum computers~\cite{IBMQ}
with the HL-VQE algorithm in Ref.~\cite{Robin:2023pgi}. 
Finally, Sec.~\ref{sec:conclu} provides a summary, conclusions and perspectives to this work.

\section{The Lipkin-Meshkov-Glick (LMG) Model and Effective-Model-Space Calculations}\label{sec:LMG}

\subsection{The LMG Model}
\begin{figure}[t]
\centering{\includegraphics[width=0.9\columnwidth] {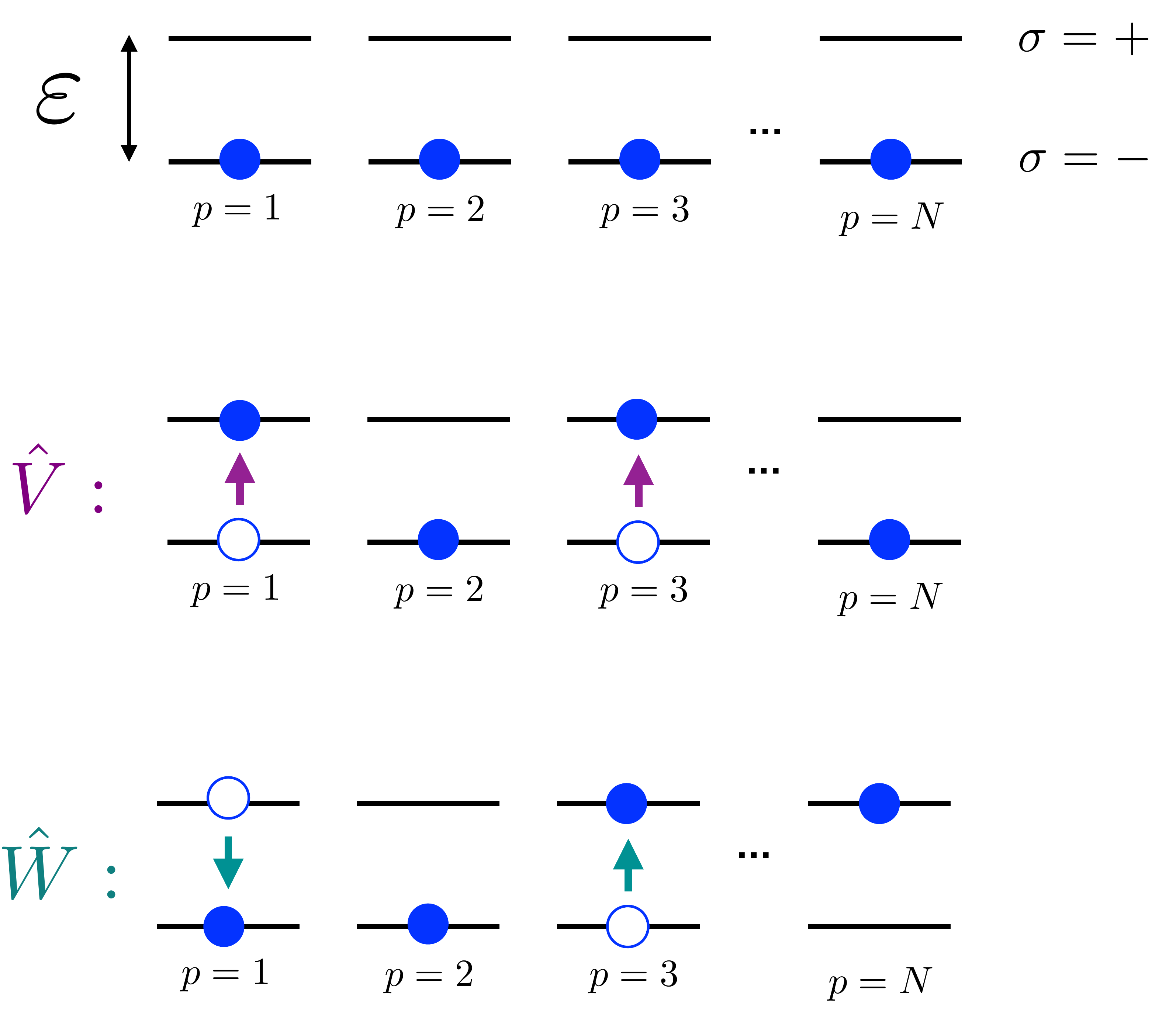}} 
\caption{Illustration of the LMG model: $N$ particles are distributed on two levels with $N$-fold degeneracy 
and interact via the terms in the Hamiltonian in Eq.~(\ref{eq:H_unrot_1}) and  (\ref{eq:H_unrot_2}) with strengths $V$ and $W$. The LMG model can be interpreted as a system of interacting spins (qubits) in an external field.}
\label{fig:LMG_levels}
\end{figure}

\noindent
In its original formulation~\cite{LIPKIN1965188}, the LMG model describes a system of $N$ identical fermions distributed on two levels $\sigma=\pm$, that are separated in energy by $\varepsilon$, each $N$-fold degenerate.
The structure and dynamics of the system are governed by the Hamiltonian 
given in Eq.~(\ref{eq:H_unrot_1}),
\begin{eqnarray}
\hat H &=& \frac{\varepsilon}{2} \sum_{\sigma p} \sigma \hat c^\dagger_{p\sigma} \hat c_{p\sigma} 
- \frac{V}{2} \sum_{p q \sigma} \hat c^\dagger_{p\sigma} \hat c^\dagger_{q\sigma} \hat c_{q-\sigma} \hat c_{p-\sigma} \nonumber \\
&&  - \frac{W}{2} \sum_{p q \sigma}  \hat c^\dagger_{p\sigma} \hat c^\dagger_{q-\sigma} \hat c_{q\sigma} \hat c_{p-\sigma} 
\; , \label{eq:H_unrot_1} \\
&=& \varepsilon \hat{J}_z 
    - \frac{V}{2} \left( \hat{J}_+^2 + \hat{J}_-^2 \right) 
    - \frac{W}{2} (\hat J_+ \hat J_- + \hat J_- \hat J_+ -\hat N) 
\; , 
\label{eq:H_unrot_2} 
 \\ \nonumber
\end{eqnarray}
where $(p,\sigma)$ denotes the single-particle states with $p=1,...,N$. 
This Hamiltonian contains a single-particle term $\varepsilon$,  
an interaction term $V$ that can scatter couples of fermions from one level to another within the same mode $p$, 
and an interaction $W$ that can excite one particle from level $\sigma=-$ to level $\sigma=+$ while de-exciting another from level $\sigma=+$ to level $\sigma=-$.
This is illustrated in Fig.~\ref{fig:LMG_levels}. 
The Hamiltonian can be further re-written as in Eq.~(\ref{eq:H_unrot_2}) 
in terms of collective (quasi)-spin operators $\hat J_z, \hat J_\pm$ 
\begin{eqnarray}
\hat{J}_z &=& \frac{1}{2} \sum_{p\sigma} \sigma \hat  c^\dagger_{p\sigma} \hat c_{p\sigma}  \; , 
\nonumber \\
\hat{J}_+ &=&  \sum_{p}  \hat  c^\dagger_{p+} \hat  c_{p-}  \; ,
\nonumber \\
\hat{J}_- &=& (J_+)^\dagger =  \sum_{p} \hat  c^\dagger_{p-} \hat  c_{p+}  \; , 
\label{eq:Js}
\end{eqnarray}
which generate an su(2) algebra.
\\
\\
\indent
Consequently, it has been realized that the LMG model can also be interpreted as 
a system of $N$ spins in $N-1$ dimensions, {\it i.e.} with all-to-all interaction of the same strength. ~\footnote{In the text, we sometimes use the words "mode", "spin" (or "qubit") interchangeably.}
The collective operators can then be expressed in terms of individuals spin operators 
$\hat \sigma_\alpha^{(p)}$ as
\begin{equation}
    \hat J_\alpha = \sum_p \hat \sigma_\alpha^{(p)}/2 \; , \hspace{1cm} \alpha = x,y,z \; .
\end{equation}
In the literature, the Hamiltonian in Eq.~(\ref{eq:H_unrot_2}) is then often re-written as
\begin{eqnarray}
 \hat{H} 
    &=& \varepsilon \hat{J}_z 
    - V_x (\hat J_x^2 + \chi \hat J_y^2) 
    + V_x \frac{1+\chi}{4} \hat N 
\; , \label{eq:H_unrot_3}
\end{eqnarray}
where the parameters $V_x$ and $\chi$ are related to $V$ and $W$ through
\begin{eqnarray}
    V = V_x \frac{1-\chi}{2} \; ,  \hspace{1cm}
    W = V_x \frac{1+\chi}{2} \; .
\end{eqnarray}
Here we used the notations of Ref.~\cite{PhysRevA.100.062104}~\footnote{ 
In the literature, the anisotropy parameter $\chi$ is also often referred to as $\gamma$.}.
For simplicity, we choose to work
with the dimensionless Hamiltonian $\hat{H}/\varepsilon$, 
which is equivalent to setting $\varepsilon=1$, and
introduce the dimensionless rescaled interaction strength $\bar{v}_x$,
\begin{eqnarray}
    \bar{v}_x = \frac{(N-1)V_x}{\varepsilon} 
    \; .
\end{eqnarray}

The Hamiltonian in Eq.~(\ref{eq:H_unrot_3}) shows that the spins are subject to an external field along the $z$ direction and interact in the $xy$ plane.
The case $V_x > 0$ corresponds to a ferromagnetic coupling, which presents a second-order phase transition at $\bar{v}_x=1$ in the mean field limit~\cite{PhysRevA.69.022107}, 
while the case $V_x < 0$ corresponds to an anti-ferromagnetic coupling with a first order phase transition at zero external field $\varepsilon =0$ \cite{PhysRevA.69.054101}.
We will focus on the former case in this work.
\\
\\
\indent
The Hamiltonian is rotationally invariant, preserves the number of fermions in a given mode $p$, and is also parity symmetric meaning that it preserves the parity (even or odd) of the number of fermions in the upper level (equivalently the number of spins pointing in the direction of the external field). Thus
\begin{eqnarray}
[\hat{H}, \hat{J}^2 ] =  0 \; ,  \hspace{0.8cm}
[\hat{H}, \hat{n}_p ] = 0 \; ,  \hspace{0.8cm}
[\hat{H}, \hat{\Pi} ] = 0 \; ,
\end{eqnarray}
where $\hat{n}_p = \sum_\sigma c^\dagger_{p\sigma} c_{p\sigma}$ counts the number of fermions in the mode $p$,
and the parity operator $\hat{\Pi}$ can be expressed in terms of the operator $\hat{N}_+$ that counts the number of particles in the upper level (number of spins up):
\begin{eqnarray}
\hat{\Pi} &=& e^{i \pi \hat{N}_+} \sim \prod_p \hat \sigma_z^{(p)} \; , \label{eq:parity_op} \\
\hat{N}_+ &=& \sum_p \hat c^\dagger_{p+} \hat c_{p+} = \hat{J_z} + \hat{N}/2 \; .
\end{eqnarray}
The exact solutions of the Hamiltonian can be conveniently expanded in the basis 
of eigenstates of
$\hat{J}^2$ and $\hat{J}_z$, $\{\ket{J,M}\}$ 
(often referred to as Dicke states),
\begin{eqnarray}
\ket{\Psi^J}_{ex} = \sum_{M=-J}^J A_{J,M} \ket{J,M} \; ,
\label{eq:wf_exact}
\end{eqnarray}
and can be straightforwardly
computed by numerical diagonalization. 
The matrix elements of the Hamiltonian and angular-momentum operators
are given in appendix~\ref{app:HME}.
In the following, 
we will restrict ourselves to the sector $J=N/2$, which contains the ground state.
It is sometimes convenient to
use the compact notation which labels the basis states as~\cite{Robin:2023pgi}
\begin{equation}
    \ket{J,M} \equiv \ket{N_+} = \ket{J= \frac{N}{2}, M = N_+ - J = N_+ - \frac{N}{2}}  
\end{equation}
where $N_+=0,...,N+1$ denotes the number of particles in the upper level (or equivalently, the number of spins up).

\subsection{Effective-Model-Space Calculations}
\begin{figure}[!ht]
\centering{\includegraphics[width=\columnwidth] {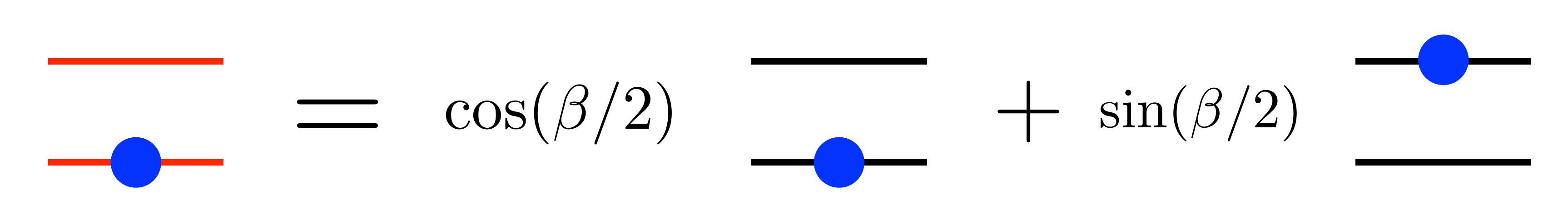}} 
\caption{ Schematic illustration of the rotation of the spins (qubits) 
by angle $\beta$ around the $y$-axis. }
\label{fig:spin_rotation}
\end{figure}

Here the goal is to apply an effective-space method, as one would apply in realistic nuclei, for which exact calculations are intractable. 
In such frameworks, 
one applies a truncation of the Hilbert space and transforms the Hamiltonian accordingly via a unitary transformation. 
This transformation is chosen so that the effective Hamiltonian acting in the truncated model 
space maximally captures the physics of the full system.
In the case of the LMG model, such truncation can be performed by introducing a cut-off, $\Lambda$, 
on the number of particles in the upper level. Here, as in Ref.~\cite{Robin:2023pgi}, we choose to perform a unitary transformation of the Hamiltonian generated by a one-body operator, which is equivalent to performing a rotation of the single-particle states (orbitals). In the LMG model this transformation takes the following simple form
\begin{eqnarray}
\begin{pmatrix}
\hat c_{p+} (\beta) \\
\hat c_{p-} (\beta)
\end{pmatrix}
= 
\begin{pmatrix}
\cos(\beta/2) & -\sin(\beta/2) \\
\sin(\beta/2) & \cos(\beta/2) \\
\end{pmatrix}
\begin{pmatrix}
\hat c_{p+} \\
\hat c_{p-}
\end{pmatrix}
\; ,
\label{eq:transfo}
\end{eqnarray}
where $\hat c_{p\sigma} \equiv \hat c_{p\sigma} (\beta=0)$ is the operator annihilating a particle in the original single-particle state $(p,\sigma)$.
Equivalently, this can be seen as a rotation of the spins around the $y$ axis by an angle $\beta$.
The transformation is depicted in Fig.~\ref{fig:spin_rotation}.

\begin{figure}[!ht]
\centering{\includegraphics[width=\columnwidth] {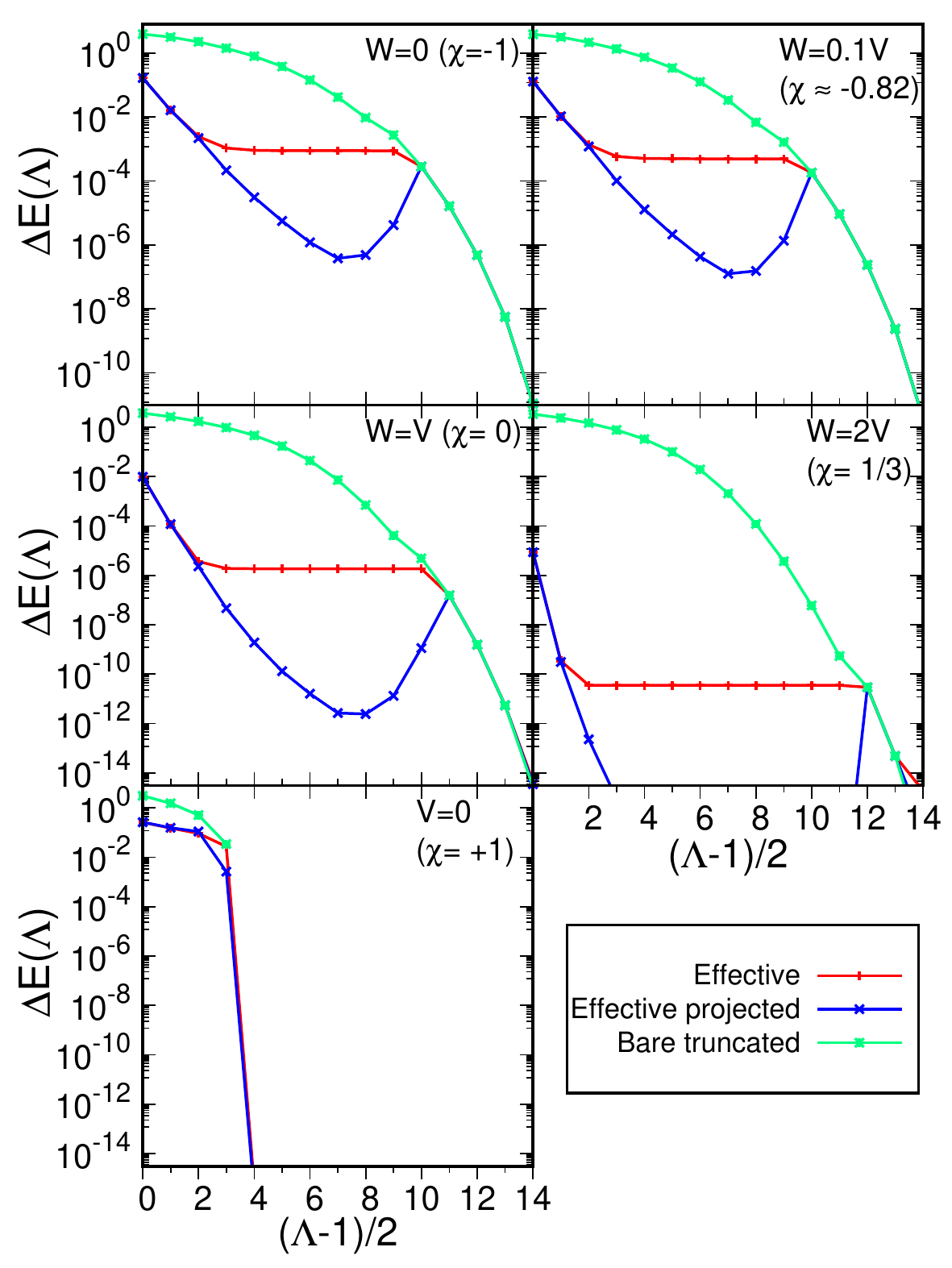}} 
\caption{Convergence of the energy error $\Delta E(\Lambda) = E(\Lambda) - E_{exact}$ as a function of the cut off $\Lambda$, for a system of $N=30$ particles, $\bar{v}_x =2.0$, and different values of $\chi$. }
\label{fig:Energy}
\end{figure}

The effective state for the system can then be written as
\begin{eqnarray}
    \ket{\Psi}^{(\Lambda)} = \sum_{N_+=0}^{\Lambda-1} A_{N_+}^{(\beta)} \ket{N_+,\beta} \; ,
    \label{eq:wf_effective}
\end{eqnarray}
where $\ket{N_+,\beta}$ are (Dicke) basis states built on the rotated 
single-particle states.
In practice, both the set of expansion coefficients $\{ A_{N_+} \}$ and the rotation angle $\beta$ are determined via energy minimization. 
The important point is that the rotation of the spins (or orbitals) does not commute with the truncation of the model space, 
and thus the effective Hamiltonian $\hat{H}(\beta)$ resulting from this transformation is expected to yield an improvement of the results, as compared to a truncation of the bare Hamiltonian 
$\hat H = \hat{H}(0)$ alone.
The full expression for the effective LMG Hamiltonian $\hat{H}(\beta)$ and the corresponding matrix elements in the basis $\{ \ket{N_+,\beta} \}$ are given in appendix~\ref{app:HME}.
\\
\\
\indent
Although in the present work we focus on the determination of the ground state, the procedure can be extended to excited states in different ways, as discussed in Ref.~\cite{Robin:2023pgi}. For example, one possibility would be to optimize the effective Hamiltonian $\hat H (\beta)$ for the ground state and consider, as an approximation, that this Hamiltonian also determines the excited states. In that case, excited states with different value of angular momentum $J$ simply correspond to minimizing the energy in a different $J$ sector, while excited states with same quantum numbers as the ground state can be obtained by adding to the effective Hamiltonian a constraint of orthogonality to the ground state $\ket{\Psi}^{(\Lambda)}$, as 
\begin{eqnarray}
    \hat H (\beta) \rightarrow \hat H' (\beta) = \hat H (\beta) - \mu \ket{\Psi}^{(\Lambda)} \, ^{(\Lambda)}\bra{\Psi} \; .
    \label{eq:new_H_constraint_orthog}
\end{eqnarray}
One can in principle repeat this procedure to determine higher-energy excited states. However, as noted in Refs.~\cite{Illa:2022zgu,Farrell:2022wyt}, this could potentially lead to an accumulation of errors in the effective Hamiltonian for highly excited states. Additionally, as observed in quantum chemistry~\cite{10.1063/1.1678164}, this procedure might fail in the case where an excited state is close in energy to the ground state but has a quite different structure, and thus requires a substantially different effective Hamiltonian. Using the ground-state Hamiltonian might thus place the excited state below the ground state.
To avoid this kind of behaviour, one could optimize the Hamiltonian parameter $\beta$ for each state using constraints of orthogonality with each previously-determined state~\cite{doi:10.1080/00268978200100722}.
As a compromise, a procedure more conventionally applied in quantum chemistry
is to optimize the orbitals via state-averaged variational
calculations~\cite{10.1063/1.1678164,https://doi.org/10.1002/qua.560200809,doi:10.1063/1.1667468}. In that case the effective Hamiltonian contains information about both ground
and excited states, and the orthonormality between states is automatically satisfied.
We leave such extensions to excited states to future work, and focus here solely on the ground state.
\\
\\
\indent
As can be intuitively inferred from Fig.~\ref{fig:spin_rotation}, the unitary transformation in Eq.~(\ref{eq:transfo}) explicitly breaks the parity symmetry associated with operator $\hat{\Pi}$ in Eq.~(\ref{eq:parity_op}), 
in the sense that the effective state can mix even and odd numbers of particles on the levels. 
Such parity symmetry should thus in principle be restored. In the present classical calculations, and in the quantum simulations performed in Ref.~\cite{Robin:2023pgi}, we chose to simply perform this symmetry restoration subsequently after variation, by projecting onto a good-parity state. More details can be found in that reference.
Other approaches, such as projection before variation, have been investigated in the context of quantum simulations, for example, in Refs.~\cite{PhysRevC.105.024324,Lacroix:2022vmg,PhysRevC.107.034310}.
\\
\\
\indent
In Ref.~\cite{Robin:2023pgi}, 
we performed effective model-space calculations for the case $\chi=-1$ $(W=0)$, 
and found that optimizing the $\beta$ angle via a variational principle allowed for an exponential speed up of the convergence of the ground-state energy and wave function with increasing cut off, 
throughout a large fraction of the Hilbert space, as opposed to the case where one performs a naive truncation of the model space with $\beta=0$ (bare Hamiltonian). 
This observation was made over a large range of values of the interaction strength $\bar{v}_x$ above (but not to close to) the phase transition.
This rapid convergence was caused by a rearrangement of the ground state components onto the basis states. In particular, it was found that the effective wave function was greatly localized around the non-interacting ground state ($N_+ = 0$) and exhibited a rapid fall-off of the components $N_+ >0$. Due to this localization, the effective state, when expressed in the original basis, was able to reproduce the exact wave function to high accuracy for low values of  $\Lambda$.
\\
\\
\indent
Below, the convergence of the results  are shown 
for values of interaction strengths other than those presented in Ref.~\cite{Robin:2023pgi}.
Specifically, 
the ratio $W/V$, corresponding to the parameter $\chi$, is varied while 
$\bar{v}_x=2.0$ is fixed.
The focus is on the convergence of the energy, as the wave function fidelity displays a similar behaviour.
Figure~\ref{fig:Energy} displays the results of our calculations, 
specifically the systematic error, $\Delta E(\Lambda) = E(\Lambda) - E_{exact}$, defined as the difference between the approximate and exact calculations, as a function of the size of the effective model space (cut off $\Lambda$). 
The red and blue curves show the results obtained with the variational transformation, 
i.e. using the effective Hamiltonian $\hat{H}(\beta)$, before and after projection, respectively.
As a comparison, results obtained from the bare Hamiltonian $\hat{H}$ with the naive truncation (green curves) are also shown.

In the case where the interaction $W<V$ (top panels), corresponding to the regime where $\chi <0$, the exponential convergence is sustained for a large part of the Hilbert space by the projection procedure. 
For large values of $\Lambda$, the model space becomes large enough so that the variational principle leads to $\beta \simeq 0$ and the transformation of the Hamiltonian becomes ineffective. 
In calculations of realistic quantum many-body systems, the portion of the Hilbert space that can be included in the model space is much smaller, so that such issues would not be encountered.
As the strength of the $W$ interaction increases, the exponential convergence accelerates, so that a high precision can be reached with smaller values of $\Lambda$. 
See, for example, the case $W=V$ ($\chi=0$) in the middle left panel.
\\
\indent 
As $W$ becomes greater than $V$ with $\chi \in (0,1)$,
it has been shown that
the exact ground state actually factorizes into a single Slater determinant when $\bar{v}_x = 1/\sqrt{\chi}$. 
This Slater determinant coincides with the Hartree-Fock mean field solution 
(corresponding to $\Lambda=1$), 
see, for example, Refs.~\cite{PhysRevA.77.052322,PhysRevA.80.062325,PhysRevA.100.062104}.
In the present case where $\bar{v}_x=2$, this corresponds to $\chi=1/4$. 
Thus, as this value is approached, the effective calculation becomes increasingly accurate for small $\Lambda$.
This can be seen, for instance, from the middle right panel ($\chi=1/3$) for which the $\Lambda=1$ 
solution reproduces the exact energy with an accuracy of $\sim 10^{-5}$, 
and which continues to improve with increasing $\Lambda$.
\\
\indent The isotropic case $\chi=1$ ($V=0$) is also special. Indeed, as can be seen from Eq.(\ref{eq:H_unrot_2}), the Hamiltonian in that case also commutes with $\hat{J}_z$ (since $\hat{J}_+\hat{J}_- + \hat{J}_-\hat{J}_+ = 2 (\hat{J}^2 - J_z)$). Thus the exact eigenstates reduce to a single $\ket{N_+}=\ket{J,M}$ state, 
where the value of $N_+$ is determined by the interaction strength $\bar{v}_x$. 
With the present value $\bar{v}_x =2.0$, $N_+$ is found to be $N_+=8$.
In the case of naively truncated calculations which use the bare Hamiltonian (green curve), if the solution $\ket{N_+}$ is included in the truncated model space, the calculation reproduces the exact result, 
but if it is outside of the model space, the result cannot recover the correct ground state, since they are orthogonal.
On the other hand, 
the effective Hamiltonian does allow a partial approach to the solution when $\Lambda < N_+$. 
Thus by rotating the operators, 
the true solution can be brought into the effective model space to some extent.
While the improvement in the energy is clearly not exponential (rather it is constant by about about one order of magnitude for low values of $\Lambda$), the effect on the entanglement entropy is much more significant,
as will be shown in the next section.
\\ \\
\indent In the general case, with $0<\chi <1$, interesting structure of the system emerges as a function of $\overline{v}_x$.
As an explicit example, we consider the case of $\chi={1\over 2}$.
For $\overline{v}_x < \sqrt{2}$, there is one ground state for the system, which in the truncated effective model spaces corresponds to a single minimum in the ground-state energy as a function of $\beta$.
With  $\overline{v}_x$ increasing from unity, the optimal value of $\beta$ evolves away from $\beta=0$, 
providing a ground-state energy that is close to the exact value.
As mentioned above, at $\overline{v}_x = \sqrt{2}$ the ground-state wavefunction is exactly 
the Hartree-Fock solution 
and therefore the effective model space minimization provides the exact ground-state energy, 
with $\beta\sim {3\over 4}$.
However, the 
minimum-energy landscape as a function of $\beta$ is broad and essentially flat, 
but with a slight monotonic positive slope.  
This must be the case for a smooth transition to the situation with $\overline{v}_x > \sqrt{2}$.
For $\overline{v}_x > \sqrt{2}$ the system has two nearly degenerate states in the ground-state manifold.
In the effective model spaces, this manifests itself as two nearly degenerate 
minima in the ground-state energy as a function of $\beta$.  Interestingly, even a three-dimensional effective model space is sufficient to identify both of these states~\footnote{However, such an energy surface does cause issues with generic energy minimization techniques.}.
\begin{figure}[!ht]
\centering{\includegraphics[width=\columnwidth]{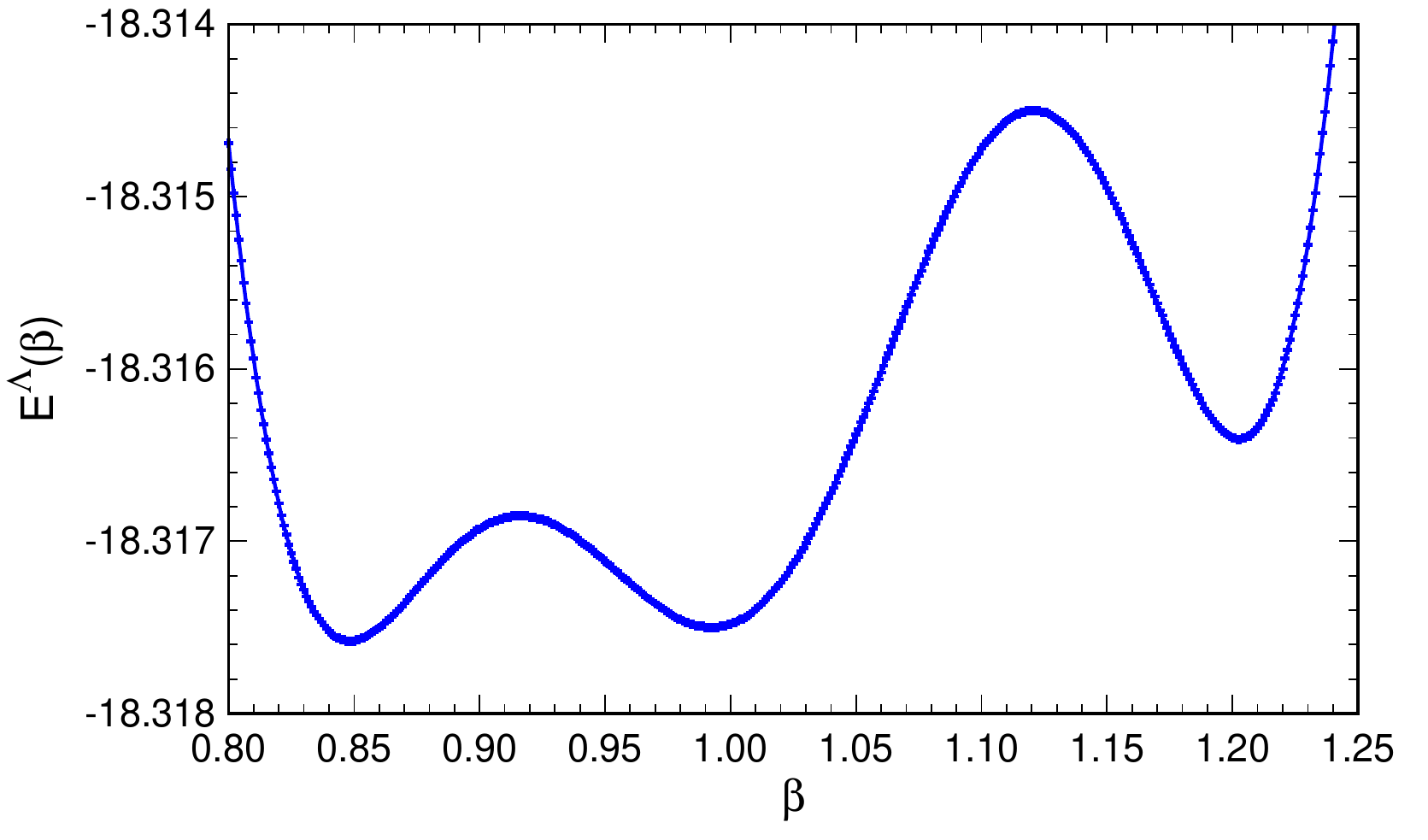}} 
\caption{The lowest eigenvalue of the effective Hamiltonian with $\Lambda=3$ for
a system with $N=30$ with $\overline{v}_x=1.92$ with $\chi={1\over 2}$ corresponding to $W/V=3$.
The two values of $\beta$ defining the two lowest lying states forming the ground-state manifold are evident.
A third minimum is also seen at larger $\beta$, but does not correspond to an actual state of the exact Hamiltonian.
}
\label{fig:betaSurface}
\end{figure}
Figure~\ref{fig:betaSurface} shows the lowest eigenvalue of the effective Hamiltonian with $\Lambda=3$ for
a system with $N=30$ with $\overline{v}_x=1.92$ with $\chi={1\over 2}$ corresponding to $W/V=3$.
This bifurcation and symmetry-breaking structures of the system as a function of $\overline{v}_x$ for a given $\chi$ is well known.

\section{Entanglement Analysis}
\label{sec:entang_LMG}

Because the LMG model represents an exactly solvable spin system, that is beyond 1D quantum spin chain, and because it presents phase transitions, the entanglement properties of this model (and possible scaling laws) have been extensively studied in the literature (see, for example, Refs.~\cite{PhysRevA.68.012101,PhysRevA.69.022107,PhysRevA.69.054101,PhysRevA.70.062304,PhysRevA.71.064101,PhysRevA.77.052105,PhysRevLett.101.025701,PhysRevA.100.062104,PhysRevB.101.054431,Calixto_2021,PhysRevA.103.032426,PhysRevA.104.032428,PhysRevA.105.062449}).
These properties have been mostly investigated in the context of exact or mean-field calculations.
We now investigate how entanglement is modified and rearranged in the effective model-space description described above, 
and compare with exact calculations that will be provided as benchmarks.
The ultimate goal is to gain insight into the present effective method, and its effect on entanglement features, to possibly guide, in the future, calculations of realistic nuclear systems.
To that aim, we will compute a few different entanglement measures. These include pure-state Von-Neumann entanglement entropies using different bi-partitionings of the system, as well as mixed-state correlation and entanglement measures such as mutual information, 2- and 4-tangles, to quantify the entanglement of different spins within a larger system.

\subsection{Von Neumann Entanglement Entropies}

In the general case of a bi-partite pure state $\ket{\Psi}_{AB} = \sum_{AB} C_{AB} \ket{\Phi}_A \otimes \ket{\Phi}_B$ describing a closed system made of two interacting sub-systems A and B, 
one set of measures of the entanglement between those two sub-systems 
are the entanglement entropies. 
Here we will consider the von Neumann entropy defined as~\cite{nielsen2010quantum} 
\begin{eqnarray}
    S_A =  - \mbox{Tr} \left( \rho_A \ln \rho_A \right) \; ,
\end{eqnarray}
where 
\begin{eqnarray}
\rho_A = \mbox{Tr}_B \ket{\Psi}_{AB\,AB}\bra{\Psi} \; ,
\label{eq:RDM_A}
\end{eqnarray}
is the reduced density of sub-system A, with $S_A=S_B$.

\begin{figure}[!ht]
\centering{\includegraphics[width=1\columnwidth] {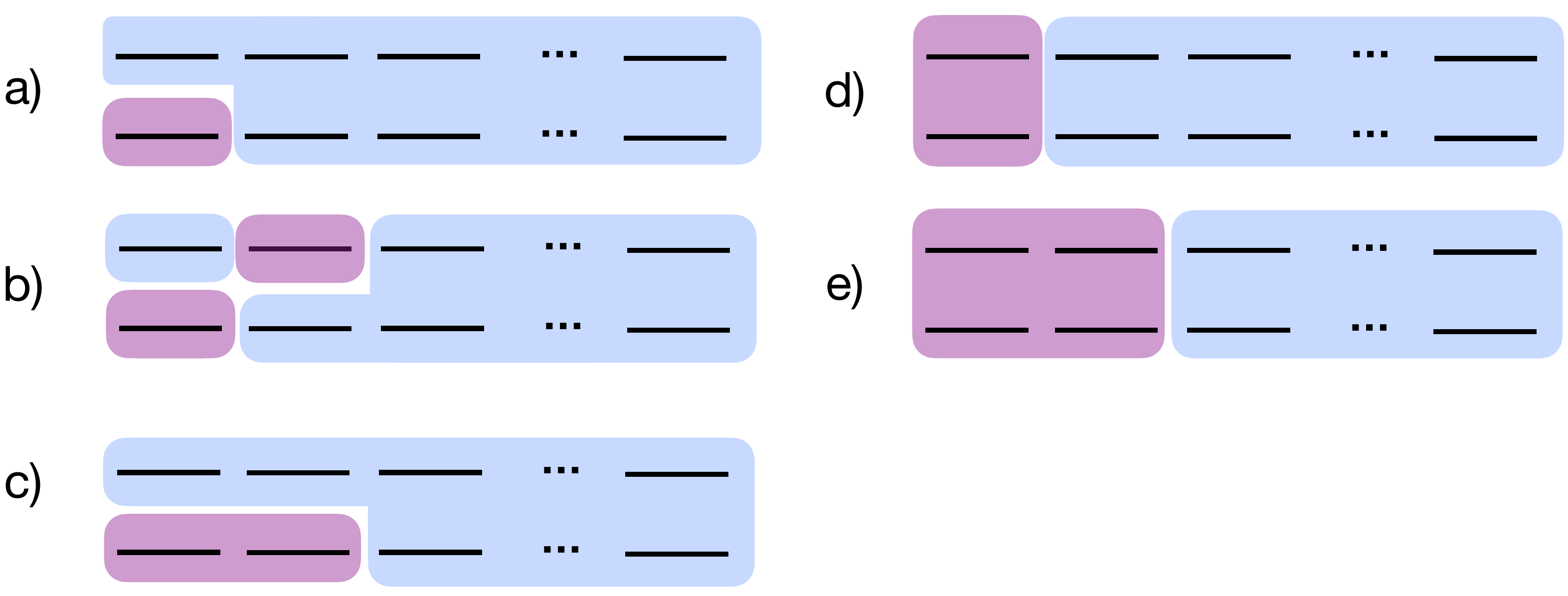}} 
\caption{Bi-partitionings of the LMG system considered in this work for calculations
of the von Neumann entanglement entropy, obtained when subsystem A (purple) is taken to be one single-particle state (orbital) (case a), two single-particle states in different modes (cases b,c), two single-particle states in the same mode which is equivalent to one spin (d), or two spins (e). }
\label{fig:bi-part}
\end{figure}

In the following we consider the bi-partitionings where subsystem A is taken to be one or two single-particle states (orbitals), as well as one or two spins, as illustrated in Fig.~\ref{fig:bi-part}.
%

\subsubsection{Orbital and Spin Entanglement Entropies}
\begin{figure}[!ht]
\centering{\includegraphics[width=\columnwidth] {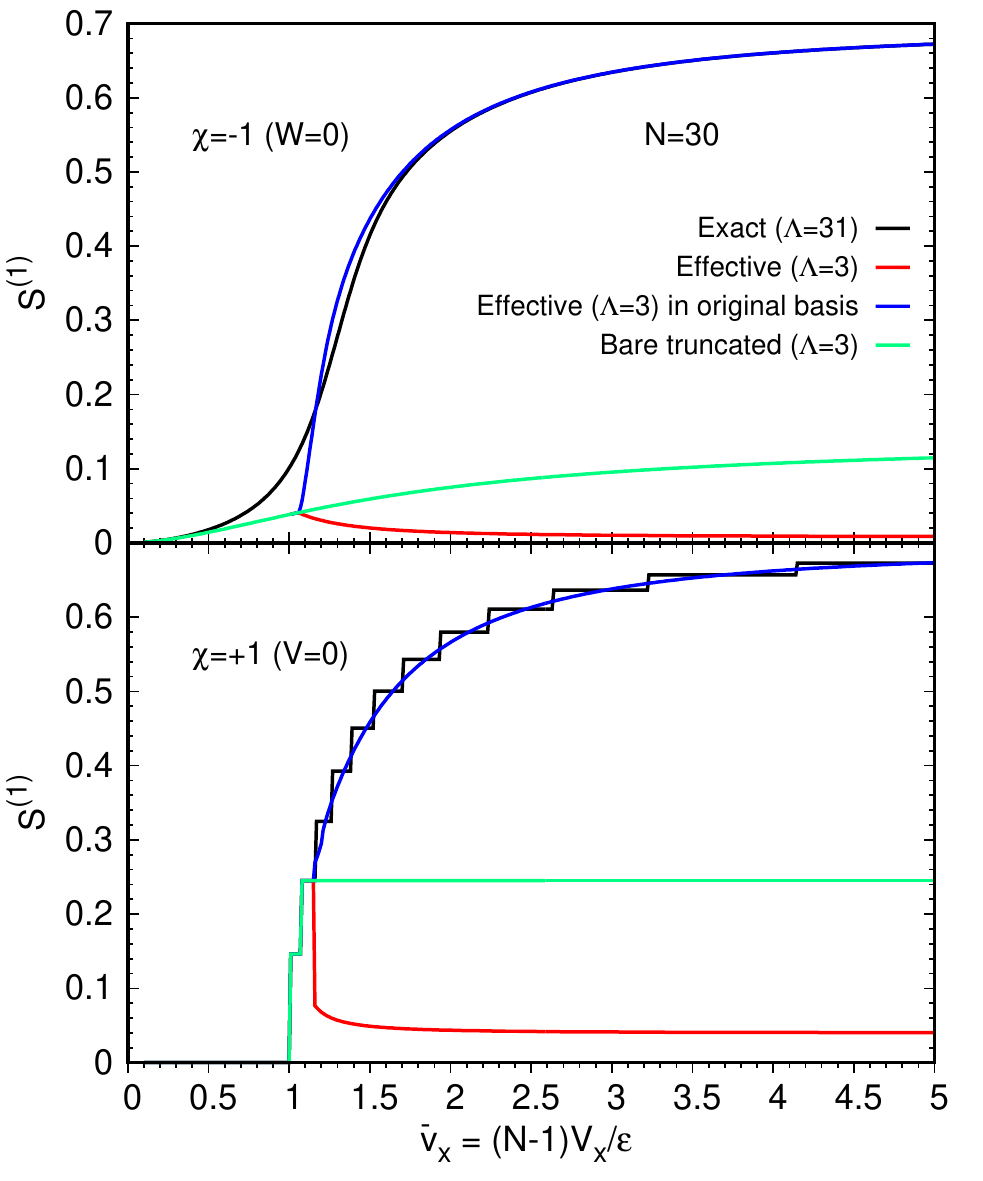}} 
\caption{One-orbital entanglement entropy as defined in Eq.~(\ref{eq:1orb_EE}) as a function of $\bar{v}_x$, for a system of $N=30$ particles, and two different values of $\chi = \pm 1 $. 
Calculations in the effective model spaces have been performed with $\Lambda=3$.}
\label{fig:1orb_EE_N30}
\end{figure}

We first consider the simplest case where A corresponds to one single-particle state (orbital) $(p,\sigma)$, as shown in Fig.~\ref{fig:bi-part}, case a).
Below we give definitions for an arbitrary wavefunction $\ket{\Psi}$ and an arbitrary set of single-particle states with creation and annihilation operators ($\hat a^\dagger_{p\sigma}$, $\hat a_{p\sigma}$). 
In practice, we will consider $\ket{\Psi}$ to be e.g. the exact or effective ground state wave function, given in Eqs.~(\ref{eq:wf_exact}) and (\ref{eq:wf_effective}) respectively, and the single-particle operators to be either the original operators $(\hat c^\dagger_{p\sigma}, \, \hat c_{p\sigma})$ or the optimized effective ones $(\hat c^\dagger_{p\sigma}(\beta), \, \hat c_{p\sigma} (\beta))$.
Since all the particles interact with each others in the same way, all modes $p=1,...N$ are equivalent in the LMG model, and thus the one-orbital reduced density matrix (RDM) is the same for all $p$. 
In the basis $\{ \ket{0} = \ket{\emptyset}, \; \ket{1} = a^\dagger_{p\sigma} \ket{\emptyset} \}$, where $\ket{\emptyset}$ denotes the particle vacuum, the one-orbital RDM is
(as derived in {\it e.g.} Ref.~\cite{Robin:2020aeh}) 
\begin{eqnarray}
\rho_{\sigma} \equiv \rho_{p,\sigma} = 
\begin{pmatrix}
1 - n_{p\sigma} & 0 \\
0 & n_{p\sigma}
\end{pmatrix}
\label{eq:1or_RDM}
\end{eqnarray}
where $n_{p\sigma}$ are occupation numbers, 
{\it e.g.}, diagonal elements of the one-body density:
\begin{eqnarray}
n_{p\pm} = \gamma_{p\pm,p\pm} &=& \langle \Psi |\hat a^\dagger_{p,\pm} \hat a_{p,\pm}| \Psi \rangle \nonumber \\
                         &=& \frac{1}{N} \langle  \Psi | \hat{N}_{\pm} | \Psi \rangle \nonumber \\
                         &=& \frac{1}{N} \langle \Psi | \hat{N}/2 \pm \hat{J}_z    | \Psi \rangle 
                         \; ,
\end{eqnarray}
where the operators $\hat N_\pm = \sum_{p} \hat a^\dagger_\pm \hat a_\pm = \hat N/2 \pm \hat J_z$ count the number of particles on each level $\sigma=\pm$, or equivalently count the number of spins up and down.
Since in the LMG model, $n_{p+} + n_{p-} =1$, 
the one-orbital entanglement entropy is the same for orbital on the upper and lower level,
\begin{eqnarray}
    S^{(1)} &=& - \left[ (1-n_{p\sigma} ) \, \mbox{ln} (1-n_{p\sigma} ) + n_{p\sigma} \,  \mbox{ln} (n_{p\sigma} )\right] \; , \forall p, \sigma. \nonumber \\
\label{eq:1orb_EE}    
\end{eqnarray}
As an example, Fig.~\ref{fig:1orb_EE_N30} displays
the one-orbital entanglement entropy for a system of $N=30$ particles as a function of $\bar{v}_x$, 
for two cases $\chi=-1,$ and $\chi = +1$.
Results obtained from exact and effective calculations with $\Lambda=3$ are shown, 
when the single-particle states are those used to expand the wave function, 
that is, the original $\beta=0$ states for the former case, 
and the optimized $\beta$ basis for the latter. 
These are shown with black and red curves, respectively.
The exact case has been calculated previously (see {\it e.g.} Ref.~\cite{PhysRevA.100.062104}), and we show it here for comparison.
In that case, the relation between phase transition and entanglement is clear, as seen from the strong increase of $S^{(1)}$ at the phase transition $\bar{v}_x \simeq 1$, which becomes maximum and saturates for large interaction strength $\bar{v}_x \gg 1$.  Thus each original orbital becomes strongly entangled with the others, at and above the phase transition. This can be taken as a signal that such orbitals are not adapted for the description of the system in this interaction regime.
In the effective framework, the entanglement entropy is drastically decreased (in fact it is minimized~\cite{PhysRevA.92.042326,Robin:2020aeh}) above the phase transition, where $\beta \ne 0$, and almost cancels at large $\bar{v}_x$ for $\chi=- 1$. 
Each single-particle state thus becomes minimally entangled with the rest of the system in the effective description. 

The fact that entanglement is lowered in principle does not necessarily mean that the effective orbitals are superior to the original ones, as small entanglement can also be an artifact of a too small model space.
This can be seen from the green curves in Fig.~\ref{fig:1orb_EE_N30}, which show the results obtained when applying a naive truncation of the model space, without optimizing the orbitals. Such simple approximation clearly leads to a poor description of the entanglement content of the system. The fact that entanglement can be artificially underestimated due to a too small model space has also been observed in realistic quantum chemistry calculations~\cite{Boguslawski_2012}. 

Thus in order to assess the quality of the effective description, meaningful comparison with the exact solution must be performed.
To do that, one can express the effective wave function 
in the original single-particle basis (characterized by $\beta=0$), as explained in Ref.~\cite{Robin:2023pgi}. 
This is shown with the blue curves in Fig.~\ref{fig:1orb_EE_N30}. 
Clearly the effective description, which includes only three many-body basis states, 
well reproduces the exact one-orbital entanglement for $\bar{v}_x \geq 1.5$.
This means that the effective wave function captures the one-body entanglement features of the exact wave function to a large extent. 
The agreement is not as good around the phase transition, which is the most difficult to capture with 
small $\Lambda$.
\\
\\
As mentioned above, the isotropic case $\chi=+1$ is special, since the exact ground state $\ket{\Psi}_{ex}$ reduces to one single basis state $\ket{N_+}=\ket{J=N/2,M}$. 
As explained in the literature (see, for example, Ref.~ \cite{PhysRevA.100.062104}), 
the transitions $N_+ \rightarrow N_+ +1$ occur as the interaction $\bar{v}_x$ increases.  
Specifically, they occur when $\bar{v}_x$ encounters integer values $\bar{v}_x^{N_+} = (N-1)/(N-2N_++ 1)$.
In particular, in the range $\bar{v}_x \leq 1$, below the phase transition, we have $N_+=0$ so that $\ket{\Psi}_{ex} = \ket{N_+=0} = \ket{J=N/2,M=-N/2}$ which reduces to a single Slater determinant where the particles all occupy the lower level, and is, by definition, unentangled (the occupation numbers $n_{p\pm}$ take values 0 or 1). 
For $\bar{v}_x >1$, the values of $N_+$ increase incrementally, up to $N_+ = N/2$ for $\bar{v}_x >> 1$. 
These increments lead to the steps in the bottom panel of Fig.~\ref{fig:1orb_EE_N30}.
In the present case where $N=30$, the value of $N_+$ increases from 0 to 1 at $\bar{v}_x = 1$, from 1 to 2 at $\bar{v}_x \simeq 1.071$, from 2 to 3 at $\bar{v}_x =1.16$ and so on.
Thus, when performing a truncation at $\Lambda=3$ using the bare Hamiltonian, the truncated solution corresponds to the exact one up to $\bar{v}_x \simeq 1.071$, above which such a naive truncation becomes clearly inadequate since the solution is orthogonal to the model space. This explains the behaviour of the green curve in the lower panel of Fig.~\ref{fig:1orb_EE_N30}. On the other hand, using the effective Hamiltonian with $\Lambda=3$ allows to capture the one-body entanglement content of the exact solution to a large extent (blue curve), while minimizing the entropy in the effective basis (red curve).
\\
\\
\indent
Similarly one can derive the two-orbital entanglement entropy, obtained from a bi-partitioning of the LMG ground state where A includes two single-particle states $(p,\sigma), (q,\sigma')$, as illustrated in case b), c) or d) of Fig.~\ref{fig:bi-part}.
In the basis \{ $\ket{0} = \ket{\emptyset}, \; \ket{1} = \hat a^\dagger_{p,\sigma} \ket{\emptyset}  , \; \ket{2} = \hat a^\dagger_{q,\sigma'} \ket{\emptyset}, \; \ket{3} =  \hat a^\dagger_{q,\sigma'} \hat a^\dagger_{p,\sigma}  \ket{\emptyset}$ \} the two-orbital RDM reads (see {\it e.g.} Ref.~\cite{Robin:2020aeh})
\begin{eqnarray}
\rho_{(ij)}=
    \begin{pmatrix}
        1-\gamma_{ii}-\gamma_{jj}+\gamma_{ijij} & 0& 0& 0\\
        0 & \gamma_{jj}-\gamma_{ijij} & \gamma_{ji} & 0\\
        0 & \gamma_{ij} & \gamma_{ii}-\gamma_{ijij} & 0\\
        0 & 0 & 0 & \gamma_{ijij}
    \end{pmatrix}
    \; , 
    \nonumber \\
\label{eq:2or-RDM}    
\end{eqnarray}
where we denote $j \equiv (p,\sigma)$ and $i \equiv (q,\sigma')$.
The one- and two-body density matrices $\gamma$ in Eq.~(\ref{eq:2or-RDM}) have the following expressions
\begin{eqnarray}
\gamma_{ij} = \gamma_{q\sigma', p\sigma}
&=&\langle \Psi| \hat a^\dagger_{p\sigma} \hat a_{q\sigma'}|\Psi\rangle \nonumber \\
&=& \left\{
\begin{array}{lll}
\langle \hat{N}_\pm \rangle /N & p=q,\sigma=\sigma\prime=\pm \\
\langle \hat{J}_\pm \rangle /N & p=q,\sigma=\pm,\sigma\prime=\mp \ \ \ \\
0       & p\ne q \\
\end{array}
\right.
\label{eq:1b_DM}
\end{eqnarray}
\begin{eqnarray}
\gamma_{ijij} 
&=& \gamma_{q\sigma',p\sigma, q\sigma',p\sigma} \nonumber \\
&=& \langle \Psi|  \hat a_{q\sigma\prime}^\dagger \hat a_{p\sigma}^\dagger \hat a_{p\sigma} \hat a_{q\sigma\prime} |\Psi\rangle \nonumber\\
&=& \left\{
\begin{array}{lll}
0 & p=q \\
\langle \hat{N}_\pm^2 \rangle/(N(N-1)) & p\ne q, \sigma=\sigma\prime=\pm \ \ \ \\
\langle \hat{N}_\pm \hat{N}_\mp\rangle/(N(N-1))     & p\ne q, \sigma=\pm, \sigma\prime=\mp \\
\end{array}
\right.
\label{eq:2b_DM}
\end{eqnarray}
where $\braket{.} \equiv \braket{\Psi|.|\Psi}$.
In deriving Eqs.~(\ref{eq:1b_DM}) and (\ref{eq:2b_DM}), the fact that all modes (spins) $p$ are equivalent has been used, so that expectation values of single-particle operators are simply expectation values of the collective ones divided by the corresponding combinatorial factor.
\\
\\
\indent
Thus for the case $p \ne q $, the 2-orbital entanglement entropy reads
\begin{eqnarray}
    S^{(2)}_{(\sigma',\sigma)} = - \mbox{Tr} \left[ \rho_{q\sigma',p\sigma} \ln (\rho_{q\sigma',p\sigma}) \right]  \; , \hspace{0.5cm} \forall (p \ne q)\; .
\end{eqnarray}
Overall, the 2-orbital entanglement entropy is found to behave in a very similar manner as the one-orbital entropy, thus we do not show it here. 
\\
\\
\indent
When $p=q$ ($\sigma' = -\sigma$) the two-orbital entanglement entropy coincides with the one-spin entanglement entropy and reduces to the following 2-dimensional matrix 
\begin{eqnarray}
\rho_{(p\sigma,p-\sigma)}= \rho_{1 \, spin}= \frac{1}{N}
\begin{pmatrix}
1-\braket{\hat N_\sigma} & \braket{\hat J_{\sigma}} \\
\braket{\hat J_{-\sigma}} & \braket{\hat N_\sigma} 
\end{pmatrix}
\; .
\label{eq:1sp_RDM}
\end{eqnarray}
When parity is preserved, and $\hat{J}_\pm = \hat{J}_\pm (\beta=0)$, the expectation value $\braket{\Psi | \hat{J}_\pm |\Psi} = 0$ so that the 2-orbital (1-spin) RDM reduces to the 1-orbital RDM given in Eq.~(\ref{eq:1or_RDM}).
In the case when the parity-broken effective wave function is used, we find that the contribution from $\braket{\Psi | \hat{J}_\pm |\Psi} = 0$ is very small. Thus in general, the one-spin entanglement entropy is equal or very close to the one-orbital entropy, as can be seen from the top panel of Fig.~\ref{fig:2spin_MI_N30}.

\subsubsection{Two-Spin Entanglement Entropy}
The two-spin entanglement entropy, corresponding to the bi-partitioning shown in Fig.~\ref{fig:bi-part} (panel e),  also coincides with the four-orbital entropy for the particular case where the orbitals are $(p,\pm)$ and $(q,\pm)$ ($p \ne q$). The two-spin entropy can thus be calculated in a number of ways, as has been done previously in the literature. 
For example, one can use a Schmidt decomposition of the $\{ \ket{N_+} =\ket{J,M} \}$ basis states~\cite{PhysRevA.71.064101,PhysRevB.101.054431} and trace out the $N-2$ remaining spins. Alternatively, one can proceed as above and derive the two-spin RDM 
$\rho_{2 \, spin}$ corresponding to the four-orbital RDM in the basis 
\{ $\ket{0} = \hat a^\dagger_{p,+} \hat a^\dagger_{q,+} \ket{\emptyset} , \; \ket{1} = \hat a^\dagger_{p,+} \hat a^\dagger_{q,-} \ket{\emptyset} , \; \ket{2} = \hat a^\dagger_{p,-} \hat a^\dagger_{q,+} \ket{\emptyset}, \; \ket{3} =  \hat a^\dagger_{p,-} \hat a^\dagger_{q,-} \ket{\emptyset}$ \} (see e.g. Ref.~\cite{PhysRevA.100.062104}), 
The matrix elements of $\rho_{2 \, spin}$ reads, $\forall p \ne q$
\begin{eqnarray}
   \rho_{00} &=&  \frac{1}{N(N-1)} \braket{ \hat N_+^2 - \hat N_+} \; ,\\
   \rho_{33} &=&  \frac{1}{N(N-1)} \braket{ \hat N_-^2 - \hat N_-} \; ,\\
   \rho_{03} &=&  \frac{1}{N(N-1)} \braket{ \hat J_-^2} \; ,\\
   \rho_{30} &=&  \frac{1}{N(N-1)} \braket{ \hat J_+^2} \; ,\\
   \rho_{11} &=& \rho_{22} = \rho_{33} = \rho_{23} = \rho_{32} = \frac{1}{N(N-1)} \braket{ \hat N_- \hat N_+}\nonumber \\
             &&  \hspace{3.1cm} = \frac{1}{N(N-1)} \braket{ \hat N_+ \hat N_-}  \; ,
\end{eqnarray}
and, 
\begin{eqnarray}
   \rho_{01} &=&  \rho_{02} = \frac{1}{N(N-1)} \braket{\hat J_+ \hat N_+} \; ,\label{eq:2spin-rdm-1} \\
   \rho_{10} &=&  \rho_{20} = \frac{1}{N(N-1)} \braket{ \hat N_+ \hat J_-}  \; ,\label{eq:2spin-rdm-2}\\
   \rho_{13} &=&  \rho_{23} = \frac{1}{N(N-1)} \braket{ \hat N_- \hat J_+}  \; ,\label{eq:2spin-rdm-3}\\
   \rho_{31} &=&  \rho_{32} = \frac{1}{N(N-1)} \braket{ \hat J_- \hat N_-} \; .\label{eq:2spin-rdm-4}
\end{eqnarray}
The matrix elements in Eqs.~(\ref{eq:2spin-rdm-1}) to (\ref{eq:2spin-rdm-4}) cancel in the case where parity symmetry is explicitly preserved. 

\begin{figure}[!ht]
\centering{\includegraphics[width=\columnwidth] {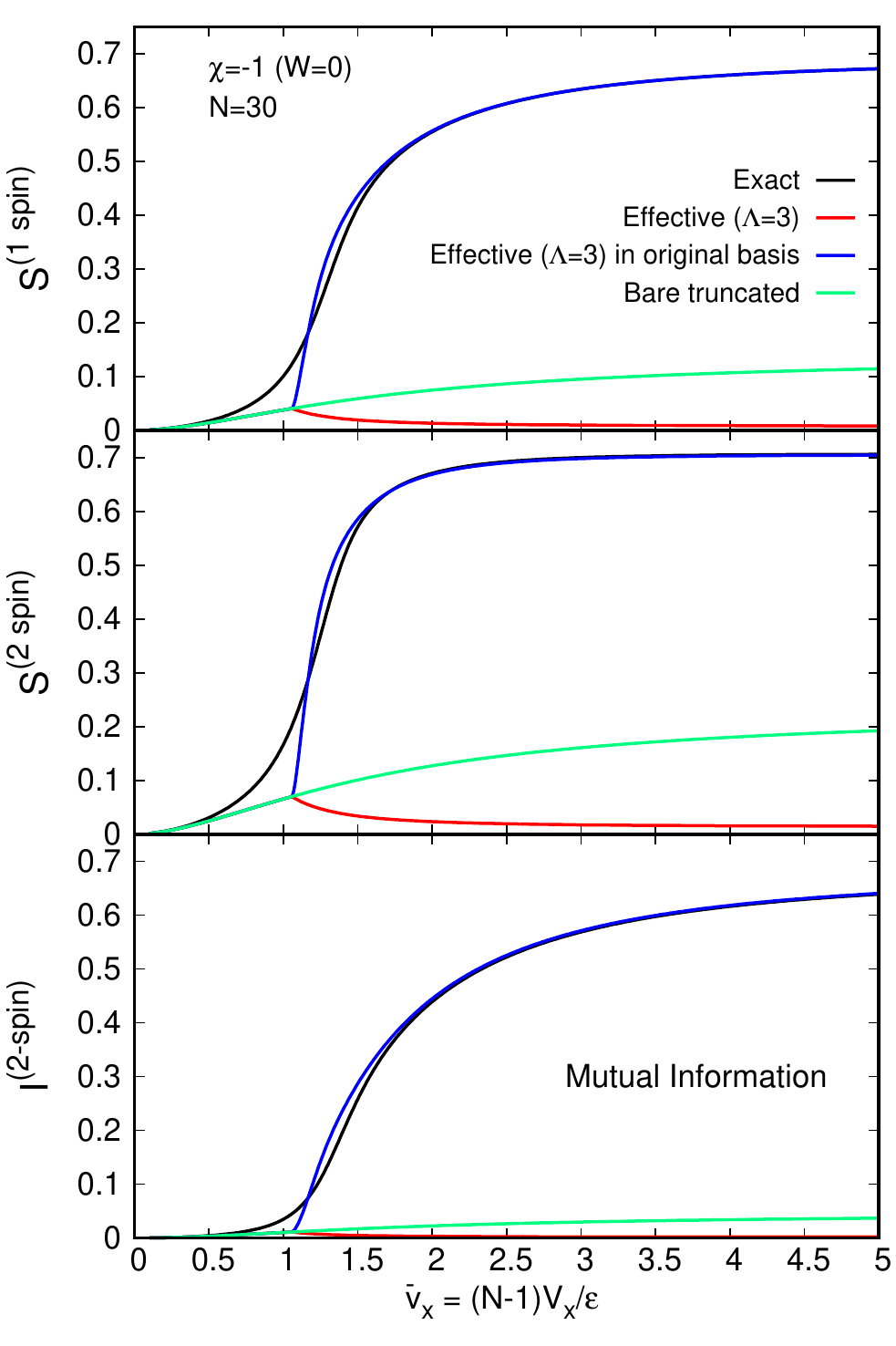}} 
\caption{One-spin (top) and two-spin (middle) entanglement entropy for $N=30$, $\chi=-1$. The two-spin mutual information is shown in the lower panel. 
}
\label{fig:2spin_MI_N30}
\end{figure}

The corresponding two-spin entanglement entropy 
\begin{eqnarray}
    S^{(2 \, spin)} = - \mbox{Tr} \left[ \rho_{2\, spin} \ln (\rho_{2\, spin}) \right]  \; , \label{eq:2-spin_VN_entropy}
\end{eqnarray}
is shown in the middle panel of Fig.~\ref{fig:2spin_MI_N30} for the case $\chi=-1$.
Compared to the one-spin entanglement entropy shown in the top panel, the two-spin entropy presents an overall similar behaviour, but a faster saturation to the maximal value in the exact calculation.

Overall the fact that the spin entanglement entropies (and mutual information, as discussed below) almost cancel in the effective description, while reproducing to a large extent the entropy of the exact state (when expressed in the original basis), is an indication that the rotated spins constitute the correct degrees of freedom for the description of the LMG system.

%

\subsection{Entanglement and Correlations Between Spins: Mutual Information and $n-$tangles}
\noindent
In the study of many-body systems, one is also interested in evaluating how a few constituents are entangled with each others, within the larger system. 
For example, one would like to characterize and quantify how A and B that are part of a larger system ABC are entangled with each others. 
In that case the density matrix for the system AB is mixed, and thus one cannot calculate the von Neumann entropy as previously described. 
Various measures of correlations and entanglement for mixed states have been introduced such as, for example, the mutual information which measures the total amount of classical and quantum correlations between two subsystems~\cite{nielsen2010quantum}, the quantum discord for the total quantum correlations (including those that are not entanglement)~\cite{PhysRevLett.88.017901}, the concurrence~\cite{PhysRevLett.80.2245}, or the $n$-tangles~\cite{PhysRevLett.78.5022,PhysRevLett.80.2245,PhysRevA.61.052306,PhysRevA.63.044301} which can quantify how $n$ constituents are entangled within the full system.

In the following we focus on quantifying entanglement between a number of spins (pairs of orbitals $(p,\pm)$), since in the LMG model, these are the relevant elementary degrees of freedom, as opposed to the individual orbitals. 
In particular we calculate the two-spin mutual information and 2-tangle, to study how two spins are entangled and classically correlated. We also investigate the $4$-tangle as a measure of the entanglement between four spins.

\subsubsection{Two-Spin Mutual Information}
A measure of the total correlations between two spins can be obtained 
by simply subtracting the von Neumann entanglement entropy of each individual spin from the two-spin entropy given in Eq.~(\ref{eq:2-spin_VN_entropy}). Thus the corresponding mutual information is defined as,
\begin{eqnarray}
    I^{(2\, spin)} = - \left[ S^{(2\, spin)} - 2 \, S^{(1\, spin)} \right] \; .
\end{eqnarray}
The results obtained within the $N=30$ system are displayed in the lower panel of Fig.~\ref{fig:2spin_MI_N30}. \\

In the exact calculation using the original ($\beta=0$) single-spin basis (black curve), the mutual information again drastically increases around and above the phase transition.
As will be shown below, the $n$-tangles display a quite different behaviour as they decrease at large $\bar{v}_x$.
This means that most of the correlations away from the phase transition are classical. 
This is in accordance with the fact that the mean field approximation usually 
provides a better description of the LMG model at large $\bar{v}_x$, 
while failing near the critical point, 
which requires a precise description of correlations that cannot be captured by a mean field.
It is clear that the effective model-space description (shown in red) is able to rearrange information, both quantum and classical, and effectively minimize all correlations existing between two spins~\footnote{
A formal proof that the present EMS  
description minimizes entanglement has, in principle, only been established for the one-body entanglement entropy~\cite{PhysRevA.92.042326,Robin:2020aeh}, 
corresponding to the (rescaled) one-orbital entropy in the case of the LMG model. 
The present results suggest that this may also be true for higher-body entanglement measures, 
in particular the two-body entropy, or the mutual information.
}.

\subsubsection{$n$-tangles} \label{sec:n-tangles}

While the mutual information provides a useful measure of correlations between constituents of a system, it does not 
provide a means to distinguish between classical and quantum contributions. 
The concurrence is a measure of entanglement that can be defined in bi-partite pure states $\ket{\Psi}_{AB}$ as $C_{AB} = 2(1-\mbox{Tr}(\rho_A^2))$ where $\rho_A$ is the reduced density matrix of susbystem A, defined in Eq.~(\ref{eq:RDM_A}).
The concurrence can also be adapted to the case when $AB$ constitute a mixed state, but only when $A$ and $B$ are single qubits, and thus can only provide a measure of two-body entanglement.
In many-body systems, one would ideally like to characterize many-particle entanglement, beyond pure two-body measures.
For that purpose, the $n$-tangle $\tau_n$ has been introduced as a measure of many-body entanglement in $N$-qubit systems with $n \leq N$~\cite{PhysRevLett.78.5022,PhysRevLett.80.2245,PhysRevA.61.052306,PhysRevA.63.044301}. It is defined as
\begin{eqnarray}
    \tau_n = |\braket{\Psi| \hat \sigma_y^{\otimes n} |\Psi^*}|^2 \; ,
    \label{eq:n_tangle_def}
\end{eqnarray}
where $\hat \sigma_y = (\hat \sigma_+ - \hat \sigma_-)/(2i)$ is the $y$-spin Pauli operator.
\\
\\
\indent
Studies of multi-partite entanglement in many-body systems of relevance for nuclear physics are in their infancy.
Beyond bi-partite entanglement, little is known about the structure of nuclei and nuclear matter.
Recently, $n$-tangles in the coherent evolution of neutrino systems
have been investigated~\cite{Illa:2022zgu,Martin:2023ljq}. 
A hierarchy $\tau_2 > \tau_3 > \tau_4.. > \tau_{N}$ was found, 
and it was revealed that $n$-body entanglement typically grows during the time evolution of a system 
initially in a tensor-product state, until reaching a plateau at large times.
It was further found that the multi-neutrino entanglement exceeded that from Bell-pairs alone, and tended toward a fixed curve with increasing system size.

In the present case, we focus on the calculation of the $n$-tangles with $n\le 4$, in order to investigate the importance of $4$-spin entanglement in the LMG interacting ground state.
In the case of a real wave function, 
i.e., $\ket{\Psi}=\ket{\Psi^*}$,
these $n$-tangles are given by
\begin{eqnarray}
    \tau_2^{(pq)} &=& |\braket{\Psi|\hat \sigma_y^p \otimes \hat \sigma_y^q |\Psi}|^2   \nonumber \\
    \tau_3^{(pqr)} &=& |\braket{\Psi|
    \hat \sigma_y^p \otimes \hat \sigma_y^q \otimes \hat \sigma_y^r|\Psi}|^2   \nonumber \\
    \tau_4^{(pqrs)} &=& |\braket{\Psi|\hat \sigma_y^p \otimes \hat \sigma_y^q 
    \otimes \hat \sigma_y^r \otimes \hat \sigma_y^s |\Psi}|^2 \; ,
\end{eqnarray}
where $\hat \sigma_y^p$ is the $y$-Pauli operator acting on spin $p$.
As detailed in appendix~\ref{app:n-tangles}, 
using the fact that the LMG model is invariant under exchanges of the spins, these $n$-tangles become,
\begin{eqnarray}
    \tau_2 \equiv \tau_2^{(pq)} 
    &=& \left| \frac{1}{N(N-1)}  \left( 4 \braket{\hat J_y^2} - N \right) \right|^2 
    \; , \ \ \forall (p \ne q) \; ,
\label{eq:2-tangle}    
\end{eqnarray}
where
\begin{eqnarray}
\braket{\hat J_y^2} 
&=& - \frac{1}{4} \left( \braket{\hat J_+^2 } +\braket{\hat J_-^2} - \braket{\hat J_+ \hat J_- + \hat J_- \hat J_+} \right) \nonumber \\ 
&=& - \frac{1}{4} \left( \braket{\hat J_+^2} +\braket{\hat J_-^2} + 2 \braket{\hat J_z^2} - N \left( \frac{N}{2} + 1 \right)  \right) 
    \; ,
\end{eqnarray}
and $\braket{.} = \braket{\Psi| . |\Psi}$.
Similarly, the 3- and 4-tangles are found to be
\begin{eqnarray}
    \tau_3 \equiv \tau_3^{(pqr)} &=& \Bigl| \frac{1}{N(N-1)(N-2)}  \Bigl[ 8 \braket{\hat J_y^3} - 2(3N-2) \braket{\hat J_y} \Bigr]  \Bigr|^2 \; , \nonumber \\
    &&\hspace{4cm} \forall (p \ne q \ne r) \; , 
\end{eqnarray}
and, 
\begin{eqnarray}
\tau_4 \equiv \tau_4^{(pqrs)} &=& \Big| \frac{1}{N(N-1)(N-2)(N-3)}  \nonumber \\
&\times &\Bigl[ 16 \braket{\hat J_y^4} - 8 (3N-4) \braket{\hat J_y^2} + 3N(N-2) \Bigr]\Big|^2 \; , \nonumber \\ \ \ &&\hspace{3cm} \forall (p\ne q \ne r \ne s) \; .
\label{eq:4-tangle}
\end{eqnarray}
Since $\braket{\hat J_y}$ and $\braket{\hat J_y^3}$ 
can be expressed in terms of expectation values of 
odd numbers of $\hat J_+$ and $\hat J_-$ operators, the 3-tangle reduces to zero when parity-symmetry is satisfied. 
Thus, both the exact many-body wave function and the parity-restored effective wave function have no three-spin entanglement.

\begin{figure}[!ht]
\centering{\includegraphics[width=\columnwidth] {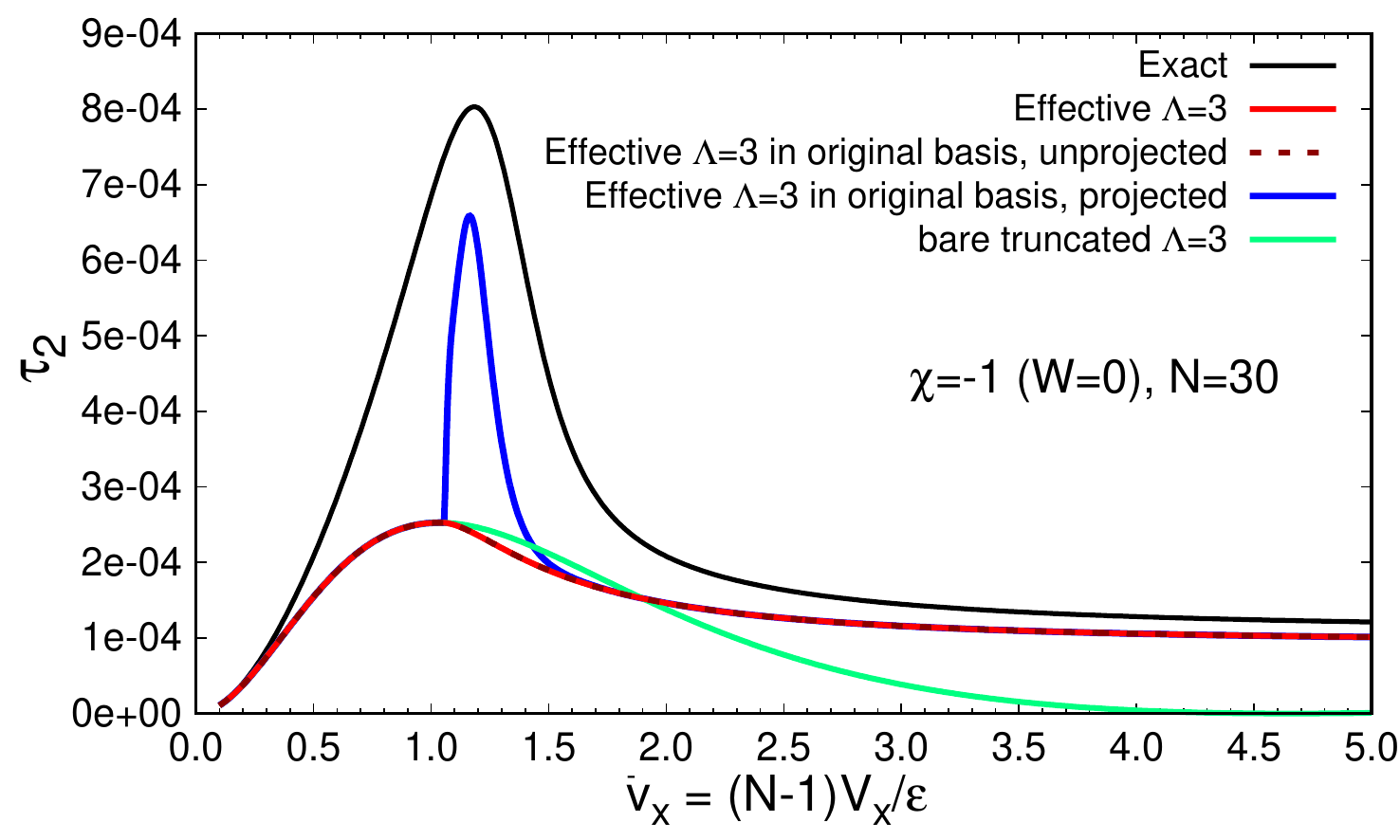}} 
\caption{ 2-tangle $\tau_2$ obtained with an EMS 
calculation with $\Lambda=3$ (red curve), and expressed in the original basis before projection (dashed red) and after projection (blue curve), for a system with $N=30$ and $\chi=-1$. The exact and truncated calculations using the bare Hamiltonian are shown in black and green, respectively.}
\label{fig:2tangle_eff}
\end{figure}

\begin{figure}[!ht]
\centering{\includegraphics[width=\columnwidth] {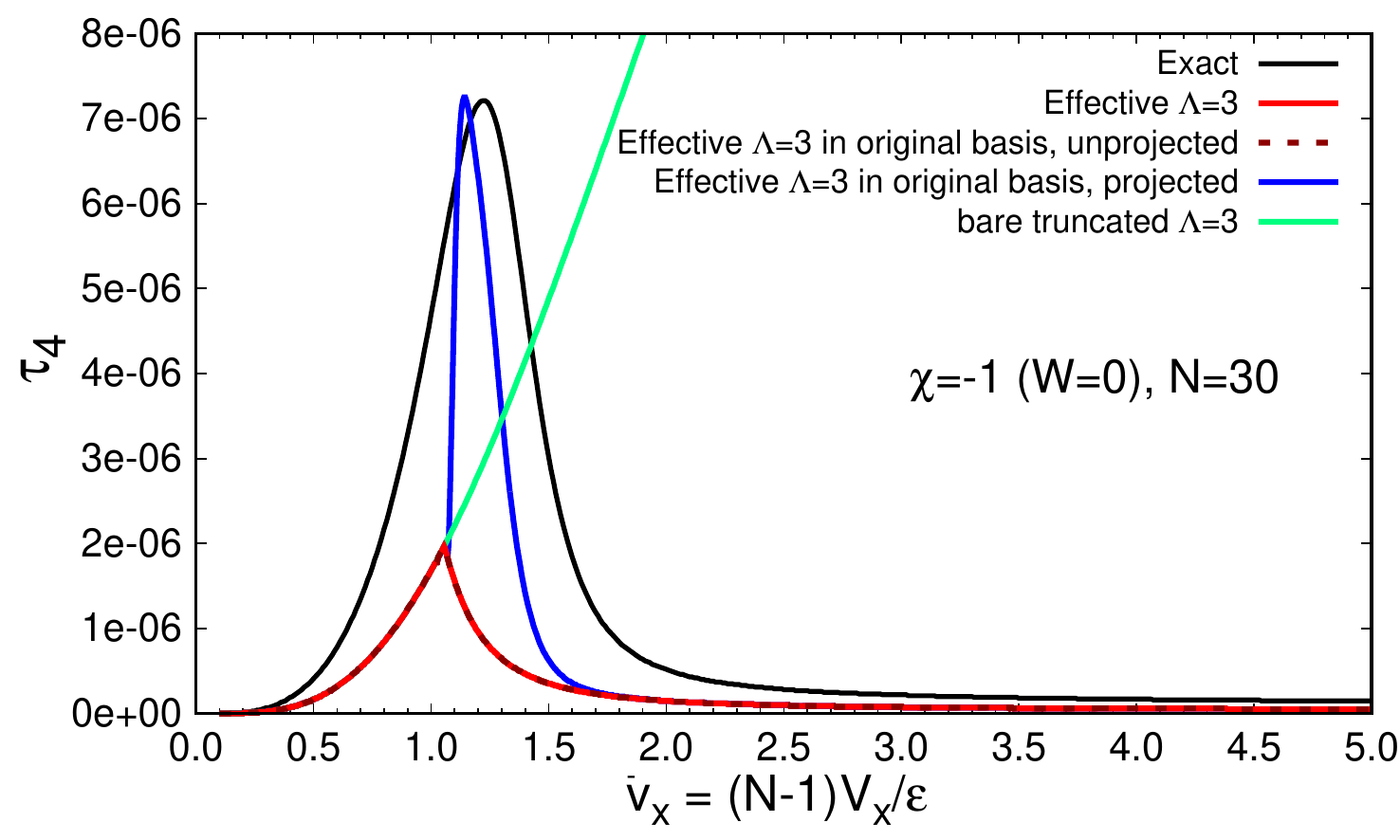}} 
\caption{4-tangle $\tau_4$ obtained with an EMS calculation with $\Lambda=3$ (red curve), and expressed in the original basis before projection (dashed red) and after projection (blue curve), 
for a system with $N=30$ and $\chi=-1$. The exact and truncated calculations using the bare Hamiltonian are shown in black and green, respectively.}
\label{fig:4tangle_eff}
\end{figure}

While we do not investigate the 5- and 6-tangles, $\tau_5 =\tau_5^{(pqrst)}$ and $\tau_6=\tau_6^{(pqrstu)}$, 
their closed form expressions are straightforward to determine,
\begin{eqnarray}
\tau_5 & = & 
\Big| {\Gamma(N-4)\over \Gamma(N+1) }\ 
\Bigl[ 
32 \braket{\hat J_y^5} - 80(N-2) \braket{\hat J_y^3} 
\nonumber \\
& &
\qquad 
+ 2 (15N^2-50N+24) \braket{\hat J_y} 
\Bigr]\Big|^2 \; , 
\nonumber \\ \ \ &&\hspace{2cm} \forall (p\ne q \ne r \ne s\ne t) \; ,
\label{eq:5-tangle}
\end{eqnarray}
and 
\begin{eqnarray}
\tau_6 & = & 
\Big| {\Gamma(N-5)\over \Gamma(N+1) }\ 
\Bigl[  64 \braket{\hat J_y^6 } - 80 (3 N-8) \braket{\hat J_y^4 }
\nonumber \\
& &  
+ 4 (45 N^2 - 210 N + 184) \braket{\hat J_y^2 }
-   15 N (N^2 - 6 N + 8)  
\Bigr]\Big|^2 \; , 
\nonumber \\ \ \ &&\hspace{2cm} 
\forall (p\ne q \ne r \ne s\ne t \ne u) \; ,
\label{eq:6-tangle}
\end{eqnarray}
where $\Gamma(k)$ is the Euler-Gamma function.
As for $\tau_3$, $\tau_5$ cancels for a parity-symmetric wave function. 
In general, $\tau_n = 0$ for odd values of $n$ and thus no odd-$n$-spin entanglement is present in the LMG model, when parity is preserved\footnote{In their original formulation, odd-$n$-tangle have been found to be undefined for $n > 3$ as they are not invariant under permutation of the qubits~\cite{PhysRevA.63.044301}. In the present LMG model, this is not an issue, since all spins are equivalent, the system is invariant under permutation of the spins, and thus, odd-$n$-tangles are in principle defined for any value of $n \le N$.
An extended definition of the odd-$n$-tangles, that eliminate this problem, has been developed in Ref.~\cite{Li_2011}.}. 
\\
\\
\indent 
Figures~\ref{fig:2tangle_eff} and \ref{fig:4tangle_eff} display $\tau_2$ and $\tau_4$ for a system of $N=30$ particles and $\chi=-1$.
In the case of exact calculations (black curves), both measures appear to be strongly peaked around the phase transition at $\bar{v}_x =1$, after which they decrease and saturate to a small value for large interaction strength $\bar{v}_x$. 
This behaviour differs from the mutual information in Fig.~\ref{fig:2spin_MI_N30} which, 
in contrast, remained at its peak value for $\bar{v}_x \gg 1$. 
This demonstrates that most of the two-spin correlations, away from the critical point, 
are classical in nature. 
Furthermore, while $\tau_2$ saturates towards a small, but non-zero value, the 4-tangle $\tau_4$ tends to cancel for $\bar{v}_x \gg 1$, pointing to the fact that in that region, 
four-spin entanglement may be safely neglected in describing the system.

The effective model-space description obtained with $\Lambda=3$ (red curves) provides a decrease of the 2- and 4-spin entanglement measures, but the effect is lesser than in previous entanglement and correlation measures. 
Results obtained after projecting onto a state of good parity are
shown with blue curves in Figs.~\ref{fig:2tangle_eff} and \ref{fig:4tangle_eff}. 
They exhibit
two distinct regions, which are separated by the critical point. Below this point, in the parity-symmetric phase, $\beta=0$ and thus the projection has no effect. 
Above the critical point, the projection improves the comparison with the exact result, however the agreement is not as good as for the previous entanglement measures.
This shows that the multi-partite quantum correlations probed by the $n$-tangles 
are more challenging to capture with effective model space calculations.

It is important to note that the $n$-tangles are, in the present LMG model, independent on the basis used to expand the wave function. This is because they only require calculations of $\braket{\hat J_y^n}$. Since the basis transformation corresponds to a rotation around the $y$ axis, which commutes with $\hat J_y$, the expectation value in Eq.~(\ref{eq:n_tangle_def}) remains invariant under such transformation. This can be appreciated by comparing the plain red and dashed dark red curves on Figs.~\ref{fig:2tangle_eff} and \ref{fig:4tangle_eff}. The $n$-tangles thus provide, in the present case, a 
basis-independent measure of multi-particle entanglement which only depends on the content of the effective many-body function. This contrasts with the previously-calculated measures, such as the entanglement entropies, which depend on both the wave function and the basis on which this wave function is expanded. 
Whether such basis-independent property of the $n$-tangles holds in many-body systems with higher-dimension elementary degrees of freedom, and more general interactions, remains to be investigated.

\begin{figure}[!ht]
\centering{\includegraphics[width=\columnwidth] {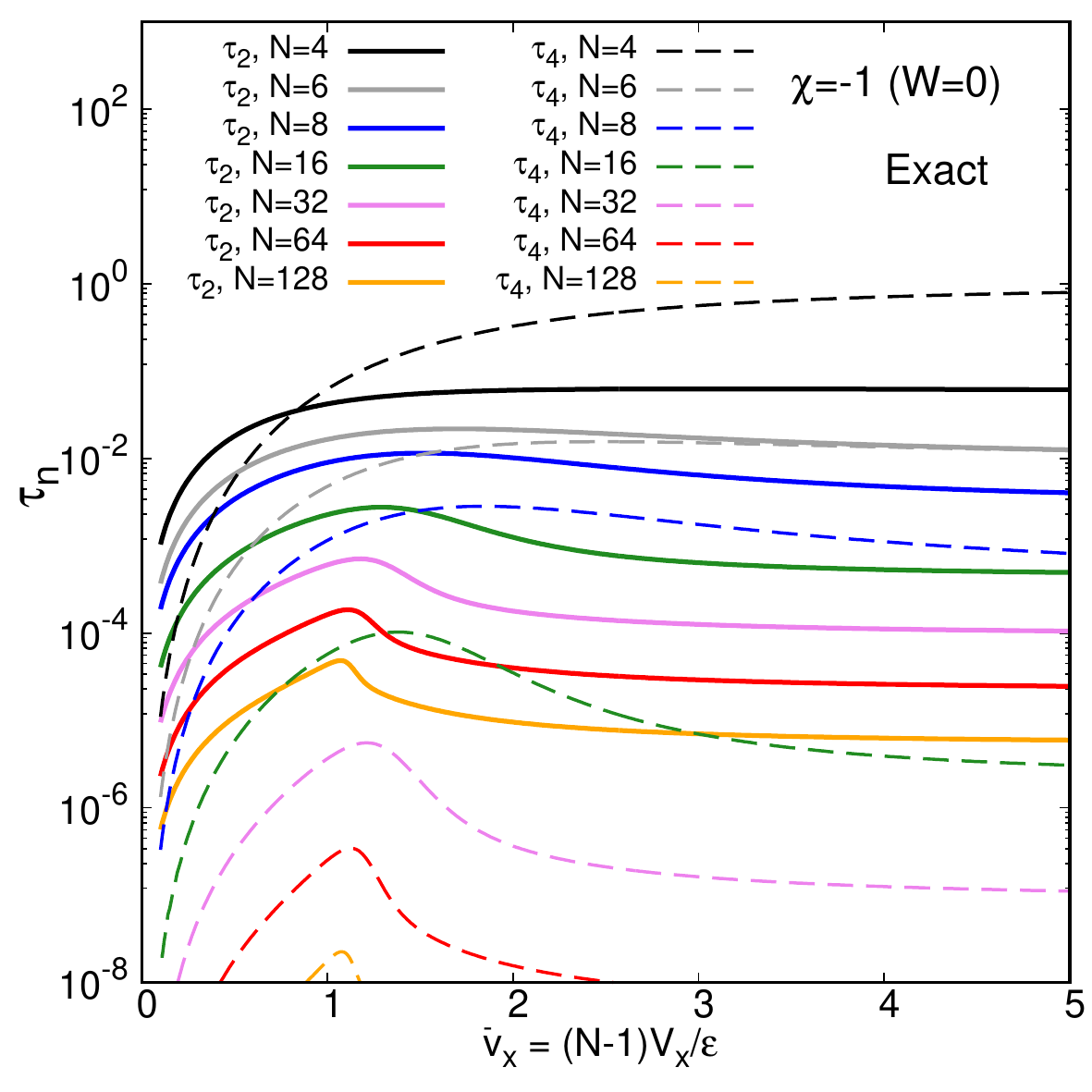}} 
\caption{The 2- and 4-tangles obtained with exact calculations of LMG systems of different sizes.
}
\label{fig:tangles}
\end{figure}

The results obtained when performing a naive truncation of
the bare Hamiltonian are displayed as the green curves in Figs.~\ref{fig:2tangle_eff} and \ref{fig:4tangle_eff}.
While the resulting $\tau_2$ exhibits the expected behaviour (based on our results above), $\tau_4$ shows a different trend than the other three curves, as it 
monotonically
increases with $\bar{v}_x$
over the displayed interval. 
Calculations for $\bar{v}_x > 5.0$ show that it saturates to a value $\simeq 4 \times 10^{-5}$.
In fact, as will be shown in the next section, truncated calculations of  
the multi-fermions quantum correlations using the bare Hamiltonian converge poorly.
\\
\\
\indent 
Studying how the $n$-tangles vary with the size of system is of prime interest, 
as this provides an estimate of the importance of $n$-particle correlations 
in the $N$-particle system.
Figure~\ref{fig:tangles} shows $\tau_2$ and $\tau_4$
as a function of $\bar{v}_x$, obtained from exact calculations with Hamiltonian parameter $\chi=-1$, for different system sizes. 
Overall, $\tau_2$ and $\tau_4$ show similar trends as for the $N=30$ system studied above.
Systems with $N \ge 8$ exhibit a $\tau_2$ that is systematically larger than the $\tau_4$
by one to several orders of magnitudes, with a difference that grows with the system size. 
Very light systems, however, appear to behave differently, as $\tau_2$ and $\tau_4$ do not decrease at large $\bar{v}_x$.
Additionally, the $N=4$ system displays a $\tau_4$ that is larger than $\tau_2$, by about one order in magnitude at large values of $\bar{v}_x$, and in the $N=6$ system $\tau_4$ is comparable to $\tau_2$.

\begin{figure}[!ht]
\centering{\includegraphics[width=\columnwidth] {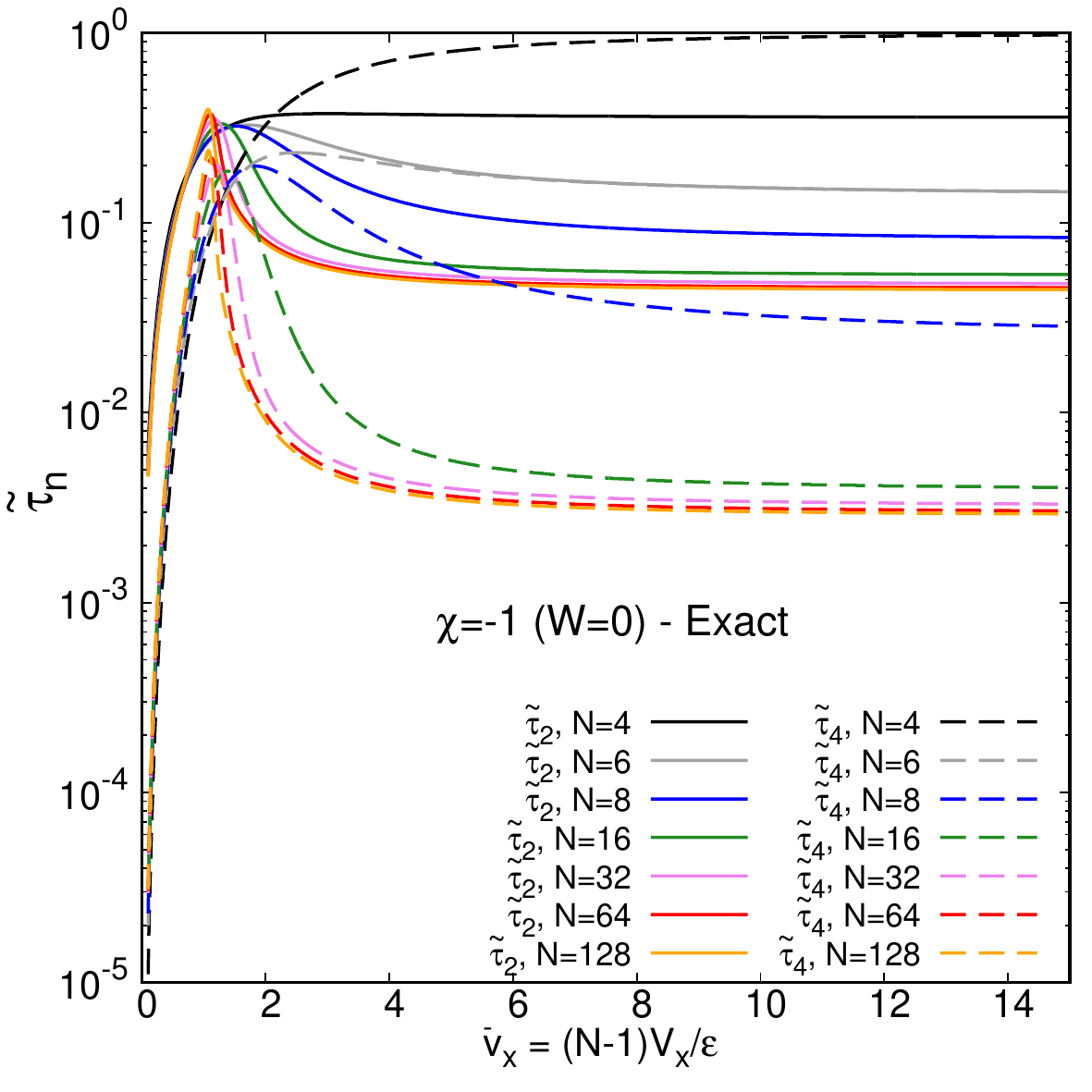}} 
\caption{The total $\widetilde{\tau}_2$ and $\widetilde{\tau}_4$ tangles, 
$\widetilde{\tau}_n = \binom{N}{n} \tau_n$,
computed in the LMG model with $\chi=-1$.}
\label{fig:tangles_rescaled}
\end{figure}

Figure~\ref{fig:tangles_rescaled} displays the total $n$-tangle $\widetilde{\tau}_n$, 
obtained by summing over all spins. In the LMG model, this simply gives
\begin{equation}
    \widetilde{\tau}_n = \binom{N}{n} \tau_n \; .
\end{equation}
It is seen that large $N$ systems converge toward the same value $\widetilde{\tau}_n$, 
independently of the interaction strength $\bar{v}_x$. 
This scaling differs from those found in
the study of neutrino systems~\cite{Illa:2022zgu}, 
which exhibit a scaling $\sim N^{n-2}$ for large $N$.
These different scalings are most likely due to the 
different forms of the Hamiltonians. 
Indeed, as seen in Eq.~(\ref{eq:H_unrot_3}), the spin-spin interaction in the LMG model is of the form (for $\chi = -1$) 
\begin{eqnarray}
    - V_x \left( \sigma^{(p)}_x \sigma^{(q)}_x - \sigma^{(p)}_y \sigma^{(q)}_y \right) \; ,
\end{eqnarray}
while the neutrino-neutrino interaction is of the form~\cite{Duan:2005cp,Balantekin_2006}
\begin{eqnarray}
    J_{pq} \left( \sigma^{(p)}_x \sigma^{(q)}_x + \sigma^{(p)}_y \sigma^{(q)}_y + \sigma^{(p)}_z \sigma^{(q)}_z\right) \; .
\end{eqnarray}
Understanding the exact relation between the form of the Hamiltonians and the large-$N$ entanglement scalings is an 
important
topic to explore, that we leave for future work.
\\
\\ \indent 
In general, it is well known that 
$\tau_2$ coincides with the squared concurrence $C_{12}^2$~\cite{PhysRevLett.78.5022,PhysRevLett.80.2245,PhysRevA.72.022309}, while 
the 3-tangle $\tau_3 = C^2_{1(23)} - (C^2_{12} + C^2_{13})$ provides a measure of genuine (irreducible) three-way entanglement. For $N \ge 4$, however, the $n$-tangles can only provide a measure of non-irreducible entanglement~\cite{li2010relationship}. Thus the fact that $\tau_4 > \tau_2$ does not necessarily mean that irreducible four-spin entanglement is larger than two-spin entanglement.

In Ref.~\cite{PhysRevA.71.042331}, the so-called $N$-qubit "residual entanglement" has been introduced as a generalization of $\tau_3$ as
\begin{eqnarray}
    \tau'_{1(2...N)} = C^2_{1(2,..,N)} - \left[ C^2_{12} + C^2_{13}+... + C^2_{1N} \right] \ge 0 \; , \nonumber \\
    \label{eq:res_entang}
\end{eqnarray}
where $C^2_{12}$ is the squared mixed-state 2-qubit concurrence, equal to the 2-tangle $\tau_2^{(12)}$~\cite{PhysRevA.72.022309}.
This measure thus subtracts the contribution of 2-way entanglement from the entanglement of one qubit (labelled as "$1$" in Eq.~(\ref{eq:res_entang})) with the rest of the system.
For general values of $N$, it is therefore not a measure of genuine irreducible $N$-way entanglement since contributions from $n$-way entanglement with $N>n>2$ can still be be present.

\begin{figure}[!ht]
\centering{\includegraphics[width=\columnwidth] {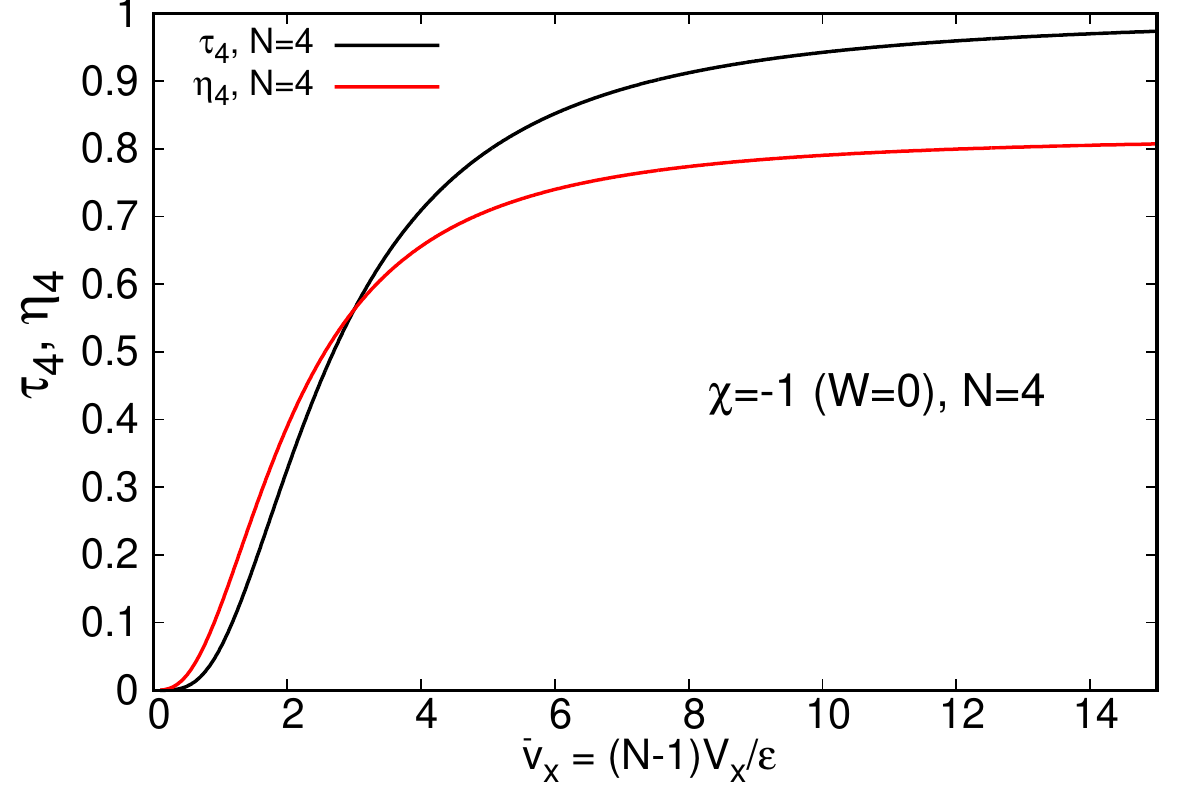}} 
\caption{"Irreducible" 4-spin entanglement $\eta_4$ in a 4-spin system, compared to the 4-tangle $\tau_4$.}
\label{fig:irr_4N_entang}
\end{figure}

For the particular case $N=4$, however, a modification of Eq.~(\ref{eq:res_entang}) that subtracts the 3-way entanglement, that is furnished by $\tau_3$, can provide a way to estimate irreducible 4-way entanglement. 
Consequently we introduce the following quantity\footnote{We note that this is entirely empirical, and whether $\eta_{1(2...N)}$ would satisfy all criteria to qualify as a proper entanglement measures has not been examined.}
\begin{eqnarray}
    \eta_{1(2...N)} &=& C^2_{1(2,..,N)} - \left[ \tau_2^{(12)} + \tau_2^{(13)}+... + \tau_2^{(1N)} \right] \nonumber \\
    &-& \left[ \tau_3^{(123)} + \tau_3^{(124)} + ... +\tau_3^{(1, N-1, N)} \right]\; .
\end{eqnarray}
In the case of the LMG model, due to invariance of the system under permutation of the spins, 
\begin{eqnarray}
    \eta_{N} = C^2_N - (N-1) \tau_2 -  \frac{(N-1)(N-2)}{2} \tau_{3} \; ,
\end{eqnarray}
where the concurrence $C^2_N = 2 (1-\mbox{Tr} (\rho_p^2)), \ \ \forall p\; ,$ 
where $\rho_p$ is the one-spin reduced density matrix given by Eq.~(\ref{eq:1sp_RDM}). 
Thus, in the case where the interacting state $\ket{\Psi}$ is parity symmetric, 
$\mbox{Tr} (\rho_p^2) = \frac{1}{2} + 2 \frac{\braket{\hat J_z}^2}{N^2}$,
from which it follows that, since $\tau_3 =0$, 
\begin{eqnarray}
    \eta_{N} = 1 -  \frac{4 \braket{\hat J_z}^2}{N^2} - 3 \tau_2 \; .
    \label{eq:irr_4N_entang}
\end{eqnarray}
In order to estimate the genuine 4-spin entanglement we compute $\eta_{4}$ for the $N=4$ system. The results are shown in Fig.~\ref{fig:irr_4N_entang}. 
It is seen that for large interaction strength, the two-spin entanglement adds significantly to the 4-body entanglement. 
\\
\\
\indent
Generalizing the procedure to estimate irreducible 4-spin entanglement in $N$-body systems with $N>4$ is difficult, because in that case the concurrence $C_{1B}$, where $B$ is a group of three spins, is not defined (since $1B$ is a mixed state of more than two qubits). One can however, define related properties, such as, for example, average concurrences as discussed in Ref.~\cite{PhysRevA.61.052306}. We plan to investigate such extensions in a future work.
Subsequently one could generalize the procedure to larger $n$, by subtracting $n-1$-way entanglement to obtain $\eta_n$. 
Overall, estimating the genuine $n$-body entanglement in $N$-body systems is of course 
difficult and such topics are being investigated in the formal context of information theory.

\subsection{The Sensitivity of Quantum Correlations to Truncations and Optimization}

\begin{figure}[!ht]
\centering{\includegraphics[width=\columnwidth] {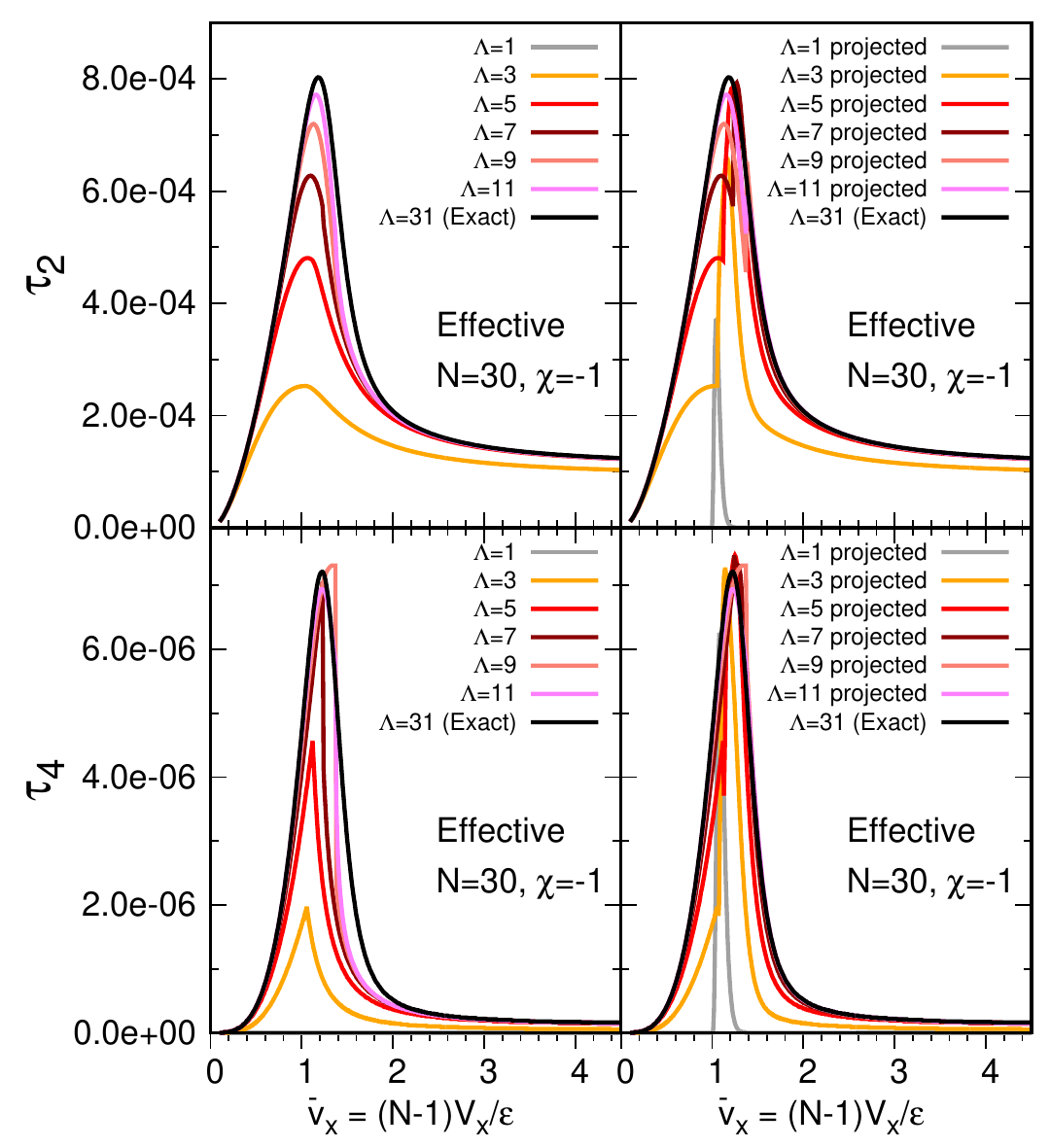}} 
\caption{Convergence of $\tau_2$ and $\tau_4$ as a function of the size of the model space $\Lambda$, obtained from EMS calculations (with $\hat{H}(\beta)$). Results obtained with the effective wave function before and after projection are displayed on the left and right panels, respectively.}
\label{fig:tangles_conv_eff}
\end{figure}

\begin{figure}[!ht]
\centering{\includegraphics[width=\columnwidth] {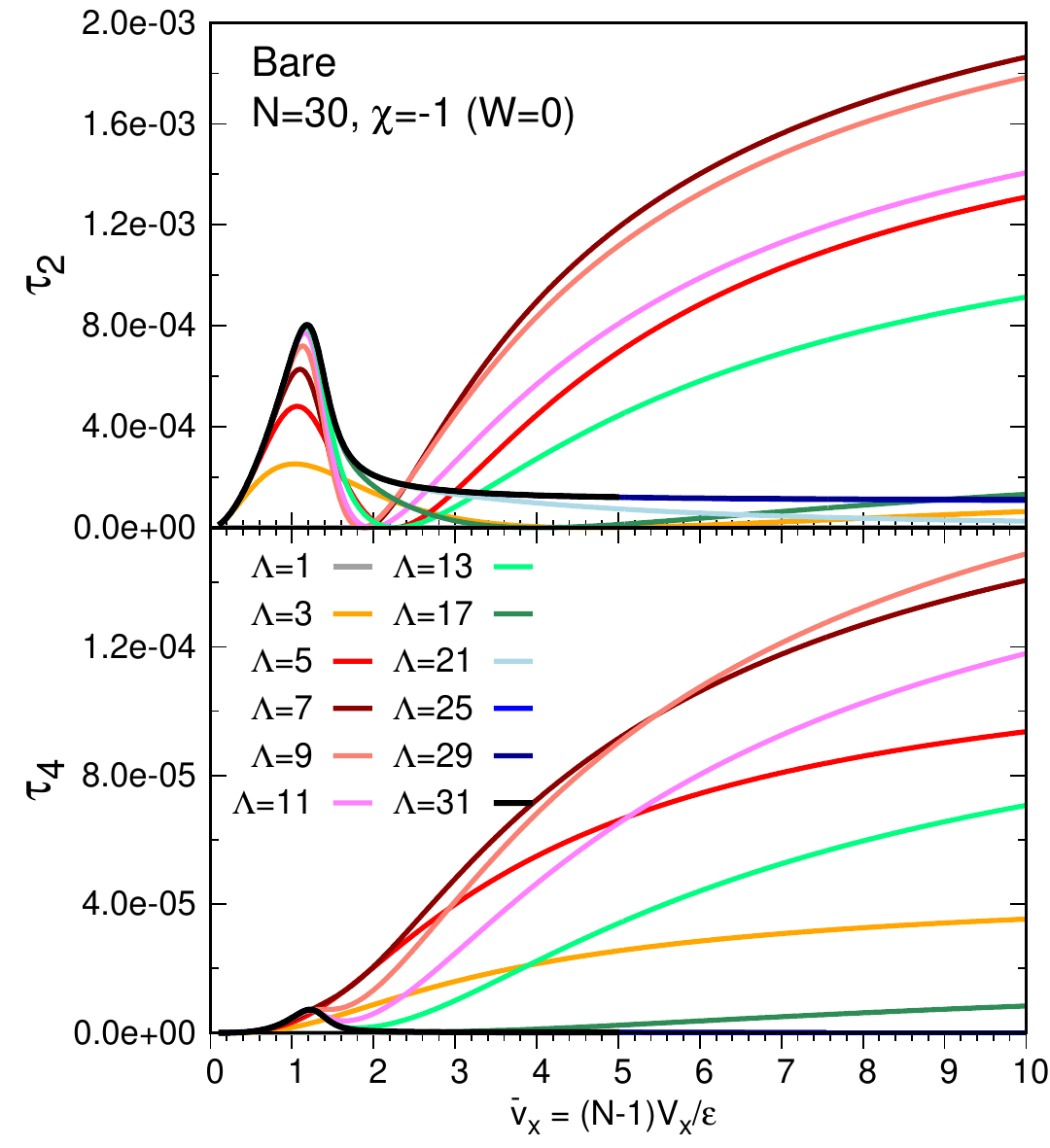}} 
\caption{Convergence of $\tau_2$ and $\tau_4$ as a function of the size of the model space $\Lambda$, obtained from naive truncations of the bare Hamiltonian $\hat{H}$.}
\label{fig:tangles_conv_bare}
\end{figure}

Calculations of realistic many-body systems, such as atomic nuclei, always involve approximations and truncations. 
Intuitively, and as confirmed by the above results, one suspects that multi-particle entanglement measures are
difficult quantities to capture in such approximate descriptions, as opposed to, for example, classical correlations or "bulk" aspects of quantum correlations, as they probe very fine details of the many-body wave function.
In this context it is interesting to study how multi-particle entanglement measures converge as the approximations are gradually lifted. 
In the present case, this can be performed by investigating the convergence of the $n$-tangles when increasing the size of the EMS, {\it i.e.} $\Lambda$. 
Below, we compare the convergence obtained with the 
EMS calculations, to the results obtained when applying a truncation of the bare 
Hamiltonian, as displayed in 
Figs.~\ref{fig:tangles_conv_eff} and \ref{fig:tangles_conv_bare}, respectively.

For the EMS calculations, Fig.~\ref{fig:tangles_conv_eff} shows results 
from the effective many-body wave functions 
before projection (left panels) and after projection (right panels).
Convergence of the results toward the exact solutions (black curves) are evident,
and improved by the projection above the phase transition (which varies with $\Lambda$).
In that region the exact results can be reproduced to a good extent with $\Lambda \gtrsim 7$.
The region below the critical point requires larger values of $\Lambda$. 
Interestingly, the 4-tangle appears to converge faster than the 2-tangle. This is counter-intuitive as one would expect higher-body correlations to converge more slowly. Whether this has a physical explanation or is simply due to the fact that $\tau_4$ exhibits a narrower peak around the phase transition is not clear.

We also note that the mean field solution ($\Lambda=1$), 
which furnishes an unentangled state
before projection, 
is largely unable to capture  
entanglement, even after projection.

Figure~\ref{fig:tangles_conv_bare} shows the convergence of $\tau_2$ and $\tau_4$ obtained 
from applying truncations of
the bare Hamiltonian, from which it is seen that 
the convergence of the multi-partite entanglement is poor, particularly in the region with large interaction strengths $\bar{v}_x \geq 2.0$. 
The convergence in that region 
does not appear to be well behaved.
Indeed, both $\tau_2$ and $\tau_4$ grow pathologically as 
$\Lambda$ increases
from $\Lambda=1$ to $\Lambda \simeq 7$, 
after which they
start decreasing again and slowly converge towards the exact value. 
In this case, 
a value $\Lambda=25$ is needed to reproduce the $\tau_2$ and $\tau_4$
of the exact calculation, which is much higher than in the 
EMS calculation.
This is a demonstration of the utility of the EMS framework.

\section{Spin squeezing}
\label{sec:spin_sq_LMG}
As the Hamiltonian of the LMG model is nonlinear in the SU(2) group generators, 
the energy eigenstates are generally spin-squeezed, 
and arbitrary tensor-product states time evolve into entangled, spin-squeezed states.
Spin squeezing~\cite{PhysRevA.47.5138,PhysRevA.50.67,PhysRevA.46.R6797}, 
entanglement and their intimate relationship in superpositions of 
$|J,M\rangle$ states, 
including the energy eigenstates of N spin-${1\over 2}$ particles (or qubits) in the LMG model, 
have been studied extensively in the context of 
quantum information~\cite{PhysRevA.68.012101}.

A number of squeezing parameters have been defined for systems with total angular momentum $J$,
\begin{eqnarray}
\xi^S & = & {2\over J}\ \left(\Delta \hat J_\perp \right)^2
\ ,
\nonumber\\
\xi^W & = &  {2\over |\langle \hat J_z \rangle| }\ \left(\Delta \hat J_\perp \right)^2
\ ,
\nonumber\\
\xi^R & = &  {2 J\over \langle \hat J_z \rangle^2 }\ \left(\Delta \hat J_\perp \right)^2
\ \ ,
\label{eq:xidefs}
\end{eqnarray}
where 
$\left(\Delta \hat J_\perp \right)^2$ 
is the minimum variance of the total angular momentum in the plane perpendicular 
to the direction of the average angular momentum of the system 
(which for this system is along the z-axis).
$\xi^S$ was defined by Kitagawa and Ueda~\cite{PhysRevA.47.5138}, 
$\xi^W$ by Wineland, Bollinger, Itano, Moore and Heinzen~\cite{PhysRevA.46.R6797},
and 
$\xi^R$ by Wineland, Bollinger, Itano and Heinzen~\cite{PhysRevA.50.67}.
Each definition of spin-squeezing is such that the system is squeezed if $\xi^i<1$.
While the most intuitive definition is that of $\xi^W$, which follows directly from the uncertainty 
principle,~\footnote{
The relevant uncertainty principle, $(\Delta \hat J_x)^2(\Delta \hat J_y)^2 \ge {1\over 4}|\langle \hat J_z\rangle|^2$,
naturally leads to a definition of squeezing $(\Delta \hat J_\perp)^2 < {1\over 2}|\langle \hat J_z\rangle|$,
and hence $\xi^W<1$.
}
$\xi^S$ seems more closely connected to entanglement~\cite{PhysRevA.68.012101}.

In our paper exploring the LMG model from the perspective of quantum simulations of the LMG model 
using effective model spaces, 
we showed that the full-space wavefunction could be systematically recovered from EMS calculations 
with exponentially improving fidelity throughout a fraction of the Hilbert space~\cite{Robin:2023pgi}.
This was enabled by the simultaneous learning of the effective Hamiltonian
and convergence to the ground-state wavefunction using the VQE algorithm (HL-VQE),
and made use of the fact that symmetries of the full Hamiltonian 
are broken by the truncation to the EMS.
With the close relation between spin-squeezing and entanglement, it is interesting to quantify the convergence 
of $\xi^2$ as a function of the truncation in the EMS.
With the most extreme truncation to an EMS with only two states ($\Lambda=2$), 
which can be simulated with a single qubit,
the ground-state wavefunction is a single eigenstate of $\hat J_z$, meaning that $A_0 =1$ and $A_1 =0$ 
in Eq.~(\ref{eq:wf_effective}).
However, when extended to four states ($\Lambda=4$), that can be simulated with two qubits, 
the ground-state becomes a superposition.

For the purposes of demonstration, 
a system with 
$N=30$ particles is explored ($J=15$), with parameters 
$\bar{v}_x = 2.0$ and $\chi=-1$,
by diagonalizing the full system (exact calculation), and by using an effective Hamiltonian in effective model spaces
with $\Lambda=2,4$ states.
HL-VQE is used to determine the $\beta$-parameter in the Hamiltonian simultaneously with the angles required to define the ground-state wavefunction using VQE.
The three spin-squeezing parameters given in Eq.~(\ref{eq:xidefs}) are shown in Table~\ref{tab:xiSWRsummary}
for the results of calculations and simulations performed in Ref.~\cite{Robin:2023pgi}.
Specifically, from the projected ground-state wavefunctions computed classically 
and from quantum simulations using IBM's 
{\tt ibm\_lagos} quantum computer, which is one of the IBM Quantum Falcon Processors~\cite{IBMQ},   
and simulator {\tt AER} using the {\tt qiskit}~\cite{Qiskit} API.
 \begin{table}[!ht]
 \centering
 \resizebox{0.45\textwidth}{!}{  
 \begin{tabular}{cccccc} 
 \hline   
\multicolumn{1}{c}{ } & \multicolumn{2}{c|}{  }& \multicolumn{2}{|c|}{ } & \multicolumn{1}{|c}{ }
\\
\multicolumn{1}{c}{ } & \multicolumn{2}{c|}{$\Lambda=2$ }& \multicolumn{2}{|c|}{$\Lambda=4$} & \multicolumn{1}{|c}{ Exact} \\
\multicolumn{1}{c}{ } & \multicolumn{2}{c|}{  }& \multicolumn{2}{|c|}{ } & \multicolumn{1}{|c}{ }
\\
 \hline
\multicolumn{1}{c}{ } & \multicolumn{1}{c}{ }& \multicolumn{1}{c}{ } & \multicolumn{1}{c}{  }& \multicolumn{1}{c}{ }
 \\
\multicolumn{1}{c}{ } & \multicolumn{1}{c}{Classical}& \multicolumn{1}{c}{{\tt ibm\_lagos}} & \multicolumn{1}{c}{Classical }& \multicolumn{1}{c}{{\tt ibm\_lagos}} & \multicolumn{1}{c}{ }
 \\
\multicolumn{1}{c}{ } & \multicolumn{1}{c}{ }& \multicolumn{1}{c}{ } & \multicolumn{1}{c}{  }& \multicolumn{1}{c}{ }
 \\
 $\xi^S$ & 1.000  & 1.019(15) & 0.6494 & 0.6670(25) & 0.5816 \\
 $\xi^W$ & 2.000  & 1.930(27)  & 1.274  & 1.311(15)   & 1.133 \\
 $\xi^R$ & 4.000  & 3.66(14)   & 2.498  & 2.579(60)  & 2.208
 \\
 \\
 \hline
 \hline
 \end{tabular}
 }
 \caption{
The spin-squeezing parameters $\xi^S$, $\xi^W$, $\xi^R$ 
of the projected ground-state wavefunction
obtained in the LMG model with parameters
$N=30$, $\varepsilon=1.0$ and $\overline{v}=2.0$
using classical simulations of the exact systems with 31 states (Exact), in effective model spaces with 
$\Lambda=2,4$ states, and using IBM's {\tt ibm\_lagos}~\cite{IBMQ}  in these spaces.
 }
 \label{tab:xiSWRsummary}
 \end{table}

The results presented in Table~\ref{tab:xiSWRsummary} show consistent convergence patterns for the three spin-squeezing parameters.
The results display (but only for the two effective model spaces we have considered, 
and for one particular set of parameters)
consistent convergence toward the full space result,
from above.  
While $\xi^S <1$ indicates that the system is spin squeezed for $\Lambda \geq 4$, the other two definitions of spin squeezing, $\xi^W$ and $\xi^R$, do not, as they are found to be greater than 1.
The ground-state wavefunctions emerging from small effective model space simulations 
are further away from being spin-squeezed than the full-space wavefunctions.

It is well-known that evolution by a Hamiltonian that is quadratic in the angular-momentum 
generators can induce squeezing from a tensor-product initial state.
For example, evolving an initial state $|J,-J\rangle_z$ forward in time via, for example,
\begin{eqnarray}
|\psi(t)\rangle & = & 
e^{-i s (\hat J_x)^2} |J,-J\rangle_z 
\ \ ,
\label{eq:tevol}
\end{eqnarray}
produces a state that is squeezed in the xy-plane, with $\xi^W < 1$ for ranges of values of $s$.
Applying this same transformation to the ground state of the LMG model with the parameters selected above, 
produces states with $\xi^W > 1$ for any value of $s$.

\section{Summary, Conclusions and Perspectives}
\label{sec:conclu}
\noindent
Better understanding the entanglement structure of quantum many-body  systems, 
including of nuclei and quantum field theories, 
and in particular how entanglement can be transformed 
to enable more efficient 
simulations of observables important to nuclear physics and high-energy physics,
is an increasingly important area of research.
Beyond the obvious intellectual-curiosity aspects, 
advances in this area will enable progress in applications, making more effective use of
classical, quantum, and hybrid
computers and associated  algorithms.
From a nuclear physics perspective, 
entanglement itself may provide measures (and insights) that can be used to develop more convergent effective descriptions of nuclei, beyond structures in the energy spectrum, adding a deeper implication to observed shell-structures.
The last few years has seen a number of groups begin to consider 
entanglement in nuclear physics settings, with a number of important and interesting results being obtained,
and a clear indication that there is considerably more to learn.
\\ \\
In this work, we have examined entanglement in the LMG model, 
which is a model used to capture elements of  nuclear many-body systems, 
but with much broader applications, including in spin systems and quantum information.
For the first time, we have considered multi-nucleon entanglement (in the context of this model),
explicitly we have derived expressions for $n$-tangles for $n\le 6$, and numerically examined the 4-tangle, $\tau_4$ for a region of parameter space.
A main focus has been to understand differences in the convergence of 
effective model space calculations of quantum correlations and entanglement versus classical correlations, 
as well as observables, such as energies.
As we have discussed previously, the non-commutivity 
of global rotations with a cut-off in the 
Hilbert space means that the Hamiltonian in any given effective model space can be variationally optimized.
Such a procedure provides effective elementary degrees of freedom which are seen to exhibit entanglement entropies and mutual information that are strongly suppressed.
The effective wavefunctions can be
transformed back into the full space, providing meaningful comparisons of convergence of the 
wavefunction and detailed correlations.
Multi-body entanglement appears to be more slowly convergent to the exact values as the dimensionality of the low-energy effective model space is increased, compared to other correlations measures and energy.
However, the rate of convergence  of the $n$-tangles  with the variationally-improved (HL-VQE) effective model space Hamiltonian is found to be even more rapid than the naive truncated space calculation using the bare, un-optimized Hamiltonian.
\\ \\
It  will be  interesting to further examine the convergence of quantum correlations in systems well-described by chiral effective field theories, that are used in realistic calculations of nuclei and nuclear many-body systems.
In particular, quantifying the quantum convergence properties of 
Weinberg's power counting~\cite{Weinberg:1990rz,Weinberg:1991um,Ordonez:1995rz},
Kaplan, Savage and Wise's (KSW) power counting~\cite{Kaplan:1998tg,Kaplan:1998we,Kaplan:2019znu}
and 
Beane, Bedaque, Savage and van Kolck's (BBSvK) power counting~\cite{Beane:2001bc}, 
for instance, will be valuable.
\\ \\
Interestingly, in the present case the $n$-tangles provide a 
basis-independent entanglement measure, as opposed to, for example, 
orbitals and spin entanglement entropies, making them of potential utility in 
probing multi-body entanglement in experiment. 
Whether such basis-independence generalizes to more general nuclear systems remains to be determined.
\\ \\
From the standpoint of quantum simulation,  the LMG model is clearly interesting in its own right, 
and also has import for quantum sensing, where the Hamiltonian and variants thereof can 
provide encoding and decoding stages of a 
multi-qubit quantum sensor.
The all-to-all connectivity of the system makes it suitable for simulation using trapped-ion systems, but presents more of a challenge with near-local connectivity, 
such as superconducting quantum computers, if using a
direct spin to qubit mapping. 
We have shown, however, that a mapping of the many-body basis states to qubits, together with the HL-VQE procedure that re-organizes information to make the wave function localized in the basis, can provide efficient simulation with superconducting quantum devices~\cite{Robin:2023pgi}. 
\\ \\
We anticipate that the results described in the present work may have applicability for
quantum simulations of lattice gauge theories (see, for example, Ref.~\cite{BDKS:2023cool}), 
such as quantum chromodynamics.
Lattice gauge theories nominally have an infinite dimensional ``bare'' state space, 
that, for example, can be truncated in SU(3) representation space.
For realistic observables, we anticipate that low-energy EMSs, suitably constructed, 
may be sufficient to perform calculations at a given level of precision, 
and further that an effective Hamiltonian in such an  EMS  can be optimized to improve convergence.
The techniques and algorithms developed for simulating nuclear many-body systems will likely
be applicable, at some level, to the quantum simulation of lattice gauge theories, 
including mean-field theory methods, 
the systematic inclusion of two-body and higher correlations,
and entanglement rearrangement.
\\ \\
To close, improving predictive capabilities for nuclear many-body systems requires a much deeper understanding of their entanglement structures and developing techniques to re-arrange and consolidate this entanglement.
This will then permit a more efficient use of quantum computing ecosystems to address grand challenge problems in nuclear physics.  We have taken steps in this direction using the LMG model.

\section*{Acknowledgments}

We would like to thank Douglas Beck, Calvin Johnson and Denis Lacroix
for helpful discussions,
and for all of our other 
colleagues and collaborators that provide the platform from which this work has emerged.
We would also like to thank the participants at the 
IQuS workshop ``At the Interface of Quantum Sensors and Quantum Simulations'', organized by
Doug Beck, Natalie Klco, Crystal Noel and Joel Ullom,
{\tt\small https://iqus.uw.edu/events/iqus-workshop-22-3b/},
for stimulating presentations and discussions.
This work was supported, in part,  
by Universit\"at Bielefeld and ERC-885281-KILONOVA Advanced Grant,
and, in part, by U.S. Department of Energy, Office of Science, Office of Nuclear Physics, Inqubator for Quantum Simulation (IQuS) under Award Number DOE (NP) Award DE-SC0020970 
via the program on Quantum Horizons: QIS Research and Innovation for Nuclear Science.
We acknowledge the use of IBM Quantum services for this work. 
The views expressed are those of the authors, 
and do not reflect the official policy or position of IBM or the IBM Quantum team.
It was supported, in part, through the Department of Physics 
\footnote{{\tt https://phys.washington.edu}}
and the College of Arts and Sciences 
\footnote{{\tt https://www.artsci.washington.edu}}
at the University of Washington. 

\clearpage
\appendix

\section{Matrix elements of the bare and effective Hamiltonians}
\label{app:HME}
\noindent
The bare Hamiltonian of the LMG model
\begin{eqnarray}
\hat H &=& \varepsilon \hat{J}_z - \frac{V}{2} \left( \hat{J}_+^2 + \hat{J}_-^2 \right) 
    - \frac{W}{2} (\hat J_+ \hat J_- + \hat J_- \hat J_+ -\hat N) \; , 
\label{eq:bare_Hami}    
\end{eqnarray}
has matrix elements in the original $\{ \ket{N_+} \equiv \ket{J,M} \}$ basis that read
\begin{eqnarray}
\braket{N_+' | \hat{H} | N_+} &=& \left( \varepsilon \, M -\frac{W}{2} \left[ C_+(M)^2 + C_-(M)^2 - N \right]\right) \delta_{N_+',N_+}  \nonumber \\
&&- \frac{V}{2} \Bigl[ C_+(M) C_+(M+1) \, \delta_{N_+',N_+ +2} \nonumber \\
&&+C_-(M) C_-(M-1) \, \delta_{N_+',N_+-2} \Bigr] \; ,
\label{eq:bare_Hami_ME}
\end{eqnarray}
where $M = N_+ - J$ and $J=\frac{N}{2}$, and where
we have defined 
\begin{eqnarray}
    C_\pm(M) = \sqrt{J(J+1) - M(M\pm 1)} \; .
\end{eqnarray}
\\
\\
\indent
The Hamiltonian acting in the effective model space can be obtained by performing a rotation $\hat{U}(\beta)$ around the $y$ axis of the collective (quasi-)spin operators $\hat{J}_\alpha$ given in Eq.~(\ref{eq:Js}), by an angle $\beta$. The rotated operators $\hat{J}_\alpha(\beta)$ are then related to the original ones 
by
\begin{eqnarray}
\hat{J}_z &=& \cos(\beta) \hat J_z(\beta)  + \frac{1}{2} \sin\beta \left( \hat J_+(\beta)  + \hat J_-(\beta)\right) \; , \\
\hat{J}_+ &=& \frac{1}{2} \Bigl[ - 2 \sin(\beta) \hat J_z(\beta)  \nonumber \\
&& + \bigl( \cos\beta +1 \bigr) \hat J_+(\beta)  + \bigl( \cos\beta -1 \bigr) \hat J_-(\beta) \Bigr] \; , \\
\hat{J}_- &=& (\hat J_+)^\dagger =  \frac{1}{2} \Bigl[ - 2\sin\beta \hat J_z(\beta)  \nonumber \\ 
&& + \bigl( \cos\beta +1 \bigr) \hat J_-(\beta) + \bigl( \cos\beta -1 \bigr) \hat J_+(\beta) \Bigr] 
\; .
\label{eq:Js_relations}
\end{eqnarray}
The resulting effective Hamiltonian then becomes
\begin{eqnarray}
\hat{H}(\beta) &=& \hat{U}^\dagger(\beta) \hat{H} \hat{U}(\beta) \nonumber \\
               &=& \hat{H}_\varepsilon(\beta) + \hat{H}_V(\beta) + \hat{H}_W(\beta) \; ,
\label{eq:eff_Hami}               
\end{eqnarray}
where 
\begin{eqnarray}
H_\varepsilon(\beta ) = \varepsilon \; \left[ \cos\beta \hat{J}_z(\beta) + \frac{1}{2} \sin\beta \bigl( \hat{J}_+(\beta) + \hat{J}_-(\beta) \bigr) \right]  \; , \nonumber \\
\end{eqnarray}
\begin{eqnarray}
\hat{H}_V(\beta) &=& - \frac{V}{4} \Bigl[ \sin^2\beta \bigl( 4 \hat{J}_z(\beta)^2 - \{ \hat{J}_+(\beta), \hat{J}_-(\beta) \} \bigr) \nonumber \\
&&  - 2 \sin\beta \cos\beta \bigl( \{ \hat{J}_z(\beta),\hat{J}_+(\beta)\} 
+ \{ \hat{J}_z(\beta), \hat{J}_-(\beta) \} \bigr) \nonumber \\
&& + (1 + \cos^2\beta) \bigl(\hat{J}_+(\beta)^2 + \hat{J}_-(\beta)^2 \bigr) \Bigr]  \; ,
\end{eqnarray}
and 
\begin{eqnarray}
\hat{H}_W(\beta) &=& - \frac{W}{4} \Bigl[  4 \sin^2\beta \hat{J}_z(\beta)^2 \nonumber \\
&&+ (1 + \cos^2\beta) \{ \hat{J}_+(\beta), \hat{J}_-(\beta) \}- 2 \hat{N} \nonumber \\
&&- 2 \sin\beta \cos\beta \bigl( \{ \hat{J}_z(\beta),\hat{J}_+(\beta)\} + \{ \hat{J}_z(\beta), \hat{J}_-(\beta) \} \bigr) \nonumber \\
&&-  \sin^2\beta \bigl(\hat{J}_+(\beta)^2 + \hat{J}_-(\beta)^2 \bigr) \Bigr] \; .\nonumber \\
\end{eqnarray}

Matrix elements of the effective Hamiltonian in Eq.~(\ref{eq:eff_Hami}) in the basis $\{ N_+, \beta\}$ can now be computed.
In contrast to the bare Hamiltonian in Eq.~(\ref{eq:bare_Hami}), the effective 
Hamiltonian can clearly couple configurations with values of $N_+$ differing by one unit, in addition to those differing by zero, or two units.
Since the rotation preserves the value of $J$, we can use $M = N_+ - J = N_+ -\frac{N}{2}$, and write the non-zero matrix elements as
\begin{eqnarray}
&&\braket{N_+, \beta |\hat{H}(\beta) | N_+, \beta} = \nonumber \\
&& \varepsilon M \cos\beta  -\frac{V}{4} \sin^2\beta \left[ 4  M^2 - [C_+(M)^2 +C_-(M)^2] \right]  \\
&&-\frac{W}{4} \left[ 4 \sin^2\beta M^2 + (1 + \cos^2\beta) [C_+(M)^2 +C_-(M)^2]  - 2N \right] 
\; , 
\nonumber
\end{eqnarray}
\begin{eqnarray}
&&\braket{N_+\pm 1, \beta |\hat{H}(\beta) | N_+, \beta} = \frac{\varepsilon}{2} \sin\beta \, C_\pm (M)   \\
&& \hspace{2cm} + \frac{V+W}{4} \left[ 2 \sin\beta \cos\beta (2M \pm 1) C_\pm(M) \right] \; , \nonumber
\end{eqnarray}
and
\begin{eqnarray}
&&\braket{N_+\pm 2, \beta |\hat{H}(\beta) | N_+, \beta} = \nonumber \\
&& \hspace{2cm} - \frac{V}{4} \left[ (1+ \cos^2 \beta) \, C_\pm (M) \, C_\pm(M\pm 1) \right] \nonumber \\
&& \hspace{2cm} + \frac{W}{4} \left[ \sin ^2 \beta \, C_\pm (M) \, C_\pm(M\pm 1) \right] \; .
\end{eqnarray}
It is straightforward to observe that the bare Hamiltonian in Eq.~(\ref{eq:bare_Hami}) and 
the matrix elements in Eq.~(\ref{eq:bare_Hami_ME}) are recovered when $\beta=0$.

\section{Derivations the n-tangles} \label{app:n-tangles}

In the case where the many-body state is real, {\it i.e.}, 
$\ket{\Psi^*} = \ket{\Psi}$, the $n$-tangles $\tau_n$ are the following,
\begin{eqnarray}
    \tau_n &=& |\braket{\Psi| \hat \sigma_y^{\otimes n} |\Psi^*}|^2 \ =\  
    |\braket{\Psi| \hat \sigma_y^{\otimes n} |\Psi}|^2 \; .
\end{eqnarray}
Their calculation thus requires the evaluation of the expectation value of the tensor products of single-spin Pauli operators $\hat \sigma_y$.
The fact that all spins are equivalent in the LMG model simplifies  
the derivation of such expectation values which can be obtained from averages of expectation values of collective operators $\hat{J_y} = \sum_p \hat \sigma_y^p/2$, where $\sigma_y^p$ acts on the single spin $p$.
\\
\\
\indent 
For example, the 2-tangle requires the calculation of
\begin{eqnarray}
    \braket{\hat \sigma_y^p \otimes \hat \sigma_y^q } 
&=& \frac{1}{N(N-1)} \sum_{p \ne q}  \braket{\hat \sigma_y^p \otimes \hat \sigma_y^q } \nonumber \\
&=& \frac{1}{N(N-1)} \Bigl[ \sum_{pq}  \braket{\hat \sigma_y^p \otimes \hat \sigma_y^q } - \sum_p \braket{(\hat \sigma_y^p )^2 } \Bigr] \nonumber \\
&=& \frac{1}{N(N-1)} \Bigl[ 4 \braket{\hat J_y^2} - N \Bigr] \; ,
\label{eq:exp_val_tau2}
\end{eqnarray}
where we used  the notation $\braket{.} = \braket{\Psi |.|\Psi}$, and the fact that $(\sigma_p)^2 =1$. One can further  insert
\begin{eqnarray}
\braket{\hat J_y^2} 
&=& - \frac{1}{4} \left( \braket{\hat J_+^2 } +\braket{\hat J_-^2} - \braket{\hat J_+ \hat J_- + \hat J_- \hat J_+} \right) \nonumber \\ 
&=& - \frac{1}{4} \left( \braket{\hat J_+^2} +\braket{\hat J_-^2} + 2 \braket{\hat J_z^2} - N \left( \frac{N}{2} + 1 \right)  \right) 
    \; ,
    \label{eq:exp_Jy2}
\end{eqnarray}
which is easily calculated in the $\{\ket{J,M} \}$ basis. 
\\
\\
\indent Similarly the expectation value for the calculation of the 3-tangle is
\begin{eqnarray}
&&    \braket{\hat \sigma_y^p \otimes \hat \sigma_y^q \otimes \hat \sigma_y^r } 
= \frac{1}{N(N-1)(N-2)} \sum_{p \ne q \ne r}  \braket{\hat \sigma_y^p \otimes \hat \sigma_y^q \otimes \hat \sigma_y^r} \nonumber \\
&& \hspace{1cm} = \frac{1}{N(N-1)(N-2)} \Bigl[ \sum_{pqr}  \braket{\hat \sigma_y^p \otimes \hat \sigma_y^q \otimes \hat \sigma_y^r} \nonumber \\
&& \hspace{1.5cm} - 3 \sum_{p \ne r} \braket{(\hat \sigma_y^p )^2 \otimes \hat \sigma_y^r} 
- \sum_p \braket{(\hat \sigma_y^p )^3 } \Bigr]  \nonumber \\
&& \hspace{1cm} = \frac{1}{N(N-1)(N-2)} \Bigl[ 8 \braket{\hat J_y^3} - 2(3N-2) \braket{\hat J_y} \Bigr] \; . \nonumber \\
\label{eq:exp_val_tau3}
\end{eqnarray}
For the 4-tangle the expectation value reads
\begin{eqnarray}
    \braket{\hat \sigma_y^p \otimes \hat \sigma_y^q \otimes \hat \sigma_y^r \otimes \hat \sigma_y^s } 
&=& \frac{1}{\widetilde{\Omega}_N} \sum_{p \ne q\ne  r \ne s} \braket{\hat \sigma_y^p \otimes \hat \sigma_y^q \otimes \hat \sigma_y^r \otimes \hat \sigma_y^s} \nonumber \\
\label{eq:exp_val_tau4}
\end{eqnarray}
where $\widetilde{\Omega}_N = N (N-1)(N-2)(N-3)$, and,

\begin{eqnarray}
&& \sum_{p \ne q\ne  r \ne s} \braket{\hat \sigma_y^p \otimes \hat \sigma_y^q \otimes \hat \sigma_y^r \otimes \hat \sigma_y^s} \nonumber \\
&=& \sum_{pqrs} \braket{\hat \sigma_y^p \otimes \hat \sigma_y^q \otimes \hat \sigma_y^r \otimes \hat \sigma_y^s} 
- 6 \sum_{p \ne r \ne s} \braket{(\hat \sigma_y^p)^2  \otimes \hat \sigma_y^r \otimes \hat \sigma_y^s} \nonumber \\
&&    - 3 \sum_{p \ne s} \braket{(\hat \sigma_y^p)^2 \otimes (\hat \sigma_y^s )^2} 
  - 4 \sum_{p \ne s} \braket{(\hat \sigma_y^p)^3 \otimes \hat \sigma_y^s} 
    - \sum_p \braket{(\hat \sigma_y^p)^4 } \nonumber \\
&=& \sum_{pqrs} \braket{\hat \sigma_y^p \otimes \hat \sigma_y^q \otimes \hat \sigma_y^r \otimes \hat \sigma_y^s} \nonumber \\
&&  \ \ \ - 2( 3N -4) \sum_{p \ne s} \braket{\hat \sigma_y^p \otimes \hat \sigma_y^s} - 3N^2 +2N
\end{eqnarray}
where 
\begin{eqnarray}
    \sum_{pqrs} \braket{\hat \sigma_y^p \otimes \hat \sigma_y^q \otimes \hat \sigma_y^r \otimes \hat \sigma_y^s} = 16 \braket{\hat J_y^4}  \; ,
\end{eqnarray}
and
\begin{eqnarray}
    \sum_{p\ne s} \braket{\hat \sigma_y^p \otimes \hat \sigma_y^s}  = 4 \braket{\hat{J}_y^2} - N  \; .
\end{eqnarray}
Thus we obtain
\begin{eqnarray}
&& \braket{\hat \sigma_y^p \otimes \hat \sigma_y^q \otimes \hat \sigma_y^r \otimes \hat \sigma_y^s } = 
 \nonumber \\
&& 
\qquad
\frac{1}{\widetilde{\Omega}_N} 
 \Bigl[ 16 \braket{\hat J_y^4} - 8 (3N-4) \braket{\hat J_y^2} + 3N(N-2) \Bigr] \; 
. \nonumber \\
\end{eqnarray}

The explicit expression for $\braket{\hat{J}_y^2}$ given in Eq.~(\ref{eq:exp_Jy2}) can be used 
to finally obtain
\begin{eqnarray}
&& \braket{\hat \sigma_y^p \otimes \hat \sigma_y^q \otimes \hat \sigma_y^r \otimes \hat \sigma_y^s } =  
\nonumber \\
&&
\frac{1}{\widetilde{\Omega}_N} 
 \Bigl[ \braket{(\hat J_+-\hat J_-)^4} + 2(3N-4) \left( \braket{\hat J_+^2} + \braket{\hat J_-^2} + 2 \braket{\hat J_z^2}\right)\nonumber \\
&& \hspace{1cm} - 3N^3 +N^2 +2N \Bigr] \; ,
\end{eqnarray}
The expectation values of products of $\hat J_\pm$ and $\hat J_z^2$ operators can again be easily calculated in the $\{\ket{J,M} \}$ basis. 

\subsection{Example of 4-tangles} 
\label{app:eg-n-tangles}
To reinforce the reader's intuition about the physical meaning of $n$-tangles,
we provide the example of $\tau_4^{abcd}$ in a system of five spins with a select wavefunction.
Consider a system with
\begin{eqnarray}
\ket{\Psi} & = & {1\over\sqrt{3}}\left[ 
|11110\rangle + |00000\rangle + |10101\rangle
\right]
\ \ \ .
\end{eqnarray}
Forming the five possible 
$\ket{\tilde\Psi}=\hat\sigma_y^a\hat\sigma_y^b\hat\sigma_y^c\hat\sigma_y^d \ket{\Psi^*}$ gives
\begin{eqnarray}
\bra{\Psi}\hat\sigma_y^1\hat\sigma_y^2\hat\sigma_y^3\hat\sigma_y^4 \ket{\Psi^*}
& = & {2\over 3}
\ \ ,\ \ 
\tau_4^{(1234)}={4\over 9}
\ \ \ ,
\end{eqnarray}
with the other four possible $\tau_4^{abcd}$ vanishing.

\subsection{Two- and four-body entanglement in the $N=4$ system} \label{app:tangle_N4}
The $N=4$ system is simple enough that it can be investigated analytically  in the general case.
Considering a (real)
wave function expanded in the $\{ \ket{N_+} = \ket{JM} \}$ basis,
as in Eq.~(\ref{eq:wf_exact}):
\begin{eqnarray}
    \ket{\Psi} = \sum_{N_+=0}^N A_{N_+} \ket{N_+} \; ,    
\end{eqnarray}
we find,
\begin{eqnarray}
\tau_2 &=& \Bigl| \frac{1}{3} A_0^2 + \frac{5}{6} A_1^2 + A_2^2 + \frac{5}{6} A_3^2 + \frac{1}{3} A_4^2 
\nonumber \\
    && -\frac{\sqrt{6}}{3} A_0 A_2 - A_1 A_3 -\frac{\sqrt{6}}{3} A_2 A_4 -\frac{1}{3} \Bigr|^2 
    \; , \nonumber \\
    & = &
    \Bigl| 
    \frac{1}{2} A_1^2 + \frac{2}{3} A_2^2 + \frac{1}{2} A_3^2  - A_1 A_3\nonumber \\
    &&\qquad  -\sqrt{\frac{2}{3}} A_0 A_2  -\sqrt{\frac{2}{3}} A_2 A_4  \Bigr|^2 \; , \\
\tau_4 &=& \left| 7 A_0^2 + 7 A_1^2 +8 A_2^2 +7 A_3^2 +7 A_4^2  -2 A_1 A_3 + 2 A_0 A_4 -7\right|^2 
\; , \nonumber \\
&=& \left| A_2^2  -2 A_1 A_3 + 2 A_0 A_4 \right|^2 
\; , \nonumber \\
\end{eqnarray}
where the norm of the wavefunction has been used to simplify the expressions.
It is clear that $\tau_2$ can be non-zero 
only if $\Lambda \ge 1$
is considered, and $\tau_4$ can be non-zero 
only if $\Lambda \ge 2$.
Subsequently, one can determine the ``irreducible'' 4-spin entanglement,
for the $N=4$ system,
given by Eq.~(\ref{eq:irr_4N_entang}) as
\begin{eqnarray}
    \eta_{4} &=& 1 -  \frac{ \braket{\hat J_z}^2}{4} - 3 \tau_2 \; , \ \
    \mbox{with} \ \ \braket{\hat J_z} = \sum_{N_+} A_{N_+} (N_+ - \frac{N}{2}) \; . \nonumber \\
\end{eqnarray}
As a pedagogical case, we consider the limit when the wave function reduces to a single 
$\ket{N_+}$ basis state.
In that case there is only one non-zero coefficient $A_{N_+} =1$.
\begin{itemize}
    \item For $N_+=0$ ($M=-2$), the state consists of a single tensor-product state with spins aligned (all particles on the lower level). Such state is unentangled and indeed it is easy to see that all $\tau_2$, $\tau_4$ and $\eta_4$ reduce to zero.
    \item For $N_+=1$ ($M= -1$), the state a W-like state, i.e. a superposition of configurations with only one of the spins up (one particle on the upper level). Such state has $\tau_2=1/9$, while  $\tau_4=\eta_4 =0$. This of course is in agreement with the well known fact that W states have only 2-body entanglement.
    \item For $N_+=2$ ($M= 0$), the state is a superposition of configurations with equal number of spins up and down (same number of particles on the upper and lower level). Such state has $\tau_2 = 1/9$, $\tau_4 =1$, and $\eta_4=2/3$, suggesting that genuine 4-spin entanglement is larger than the 2-spin entanglement. 
\end{itemize}
A state with $M>0$ has same entanglement properties as that with $M<0$.
The general case when the wave function is a superposition of states $\ket{N_+}$ depends on the values of the coefficients which are governed by the interaction of the LMG model. 
The numerical results are shown in Sec.~\ref{sec:n-tangles}.

\bibliographystyle{sn-aps}
\bibliography{biblio}
\end{document}